\documentclass[table,fleqn,usenatbib]{mnras}

\usepackage{newtxtext,newtxmath}

\usepackage[T1]{fontenc}
\usepackage{ae,aecompl}

\usepackage{graphicx}	
\usepackage{amsmath}	
\usepackage{layouts}
\usepackage{subfigure}
\usepackage{pdflscape}
\usepackage{lineno}
\usepackage[final]{changes}
\usepackage{comment}

\newcommand{\gaia}{\textit{Gaia}}
\newcommand{\LISA}{\textit{LISA}}
\newcommand{\LSST}{Vera C. Rubin observatory}
\newcommand*\rot{\rotatebox{-55}}
\newcommand*\gb{\cellcolor{gray!10}}

\newcommand{\unsim}{\mathord{\sim}} 
\newcommand{\Msun}{$\mathrm{M_{\sun}}$}
\newcommand{\Rsun}{$\mathrm{R_{\sun}}$}
\newcommand{\logg}{$\log{g}$}
\newcommand{\Teff}{$T_\mathrm{eff}$}
\newcommand{\A}{ZTFJ1637+49}	
\newcommand{\B}{ZTFJ0220+21}	
\newcommand{\C}{ZTFJ2252\textminus05}	
\newcommand{\D}{ZTFJ0407\textminus00}	
\newcommand{\E}{ZTFJ0003+14}	

\title[ZTF eclipsing AM~CVn]{Discovery and characterization of five new eclipsing AM~CVn systems}

\author[J.~van~Roestel et al.]{J.~van~Roestel,$^{1}$\thanks{E-mail: jvanroes@caltech.edu}
T. Kupfer,$^{2}$
M.~J. Green,$^{3}$
S. Wong,$^{4}$
L. Bildsten,$^{4,5}$
\newauthor
K. Burdge,$^{1}$ 
T. Prince,$^{1}$
T.~R. Marsh,$^{6}$
P. Szkody,$^{7}$
\newauthor
C. Fremling,$^{1}$ 
M.~J. Graham,$^{1}$
V.~S. Dhillon,$^{8,9}$
S.~P. Littlefair,$^{8}$
\newauthor
E.~C. Bellm$^{10}$,
M. Coughlin,$^{11}$
D.~A. Duev$^{1}$,
D.~A.~Goldstein$^{12}$
R.~R. Laher,$^{13}$
\newauthor
B. Rusholme,$^{13}$
R. Riddle,$^{14}$
R. Dekany,$^{1}$
S.~R. Kulkarni$^{1}$
\newauthor
\\
$^{1}$Division of Physics, Mathematics, and Astronomy, California Institute of Technology, Pasadena, CA 91125, USA\\
$^{2}$Department of Physics \& Astronomy, Texas Tech University, Box 41051, Lubbock TX 79409-1051, USA\\
$^{3}$Department of Astrophysics, Faculty of Exact Sciences, Tel-Aviv University, Ramat Aviv, Tel-Aviv 6139001, Israel\\
$^{4}$Department of Physics, University of California, Santa Barbara, CA 93106, USA\\
$^{5}$Kavli Institute for Theoretical Physics, University of California, Santa Barbara, CA 93106, USA\\
$^{6}$Department of Physics, University of Warwick, Coventry CV4 7AL, UK\\
$^{7}$Department of Astronomy, University of Washington, Seattle, WA 98195, USA\\
$^{8}$Department of Physics and Astronomy, University of Sheffield, Sheffield S3 7RH, UK\\
$^{8}$Instituto de Astrofisica de Canarias, E-38205 La Laguna, Tenerife, Spain\\
$^{9}$DIRAC Institute, Department of Astronomy, University of Washington, 3910 15th Avenue NE, Seattle, WA 98195, USA\\
$^{10}$School of Physics and Astronomy, University of Minnesota, Minneapolis, Minnesota 55455, USA\\
$^{11}$Weights \& Biases, Inc., 255 Kansas St, San Francisco, CA 94103, USA\\
$^{12}$IPAC, California Institute of Technology, 1200 E. California Blvd, Pasadena, CA 91125, USA\\
$^{13}$Caltech Optical Observatories, California Institute of Technology, Pasadena, CA  91125, USA\\
}

\date{Accepted XXX. Received YYY; in original form ZZZ}

\pubyear{2021}

\begin{document}

\label{firstpage}
\pagerange{\pageref{firstpage}--\pageref{lastpage}}
\maketitle

\begin{abstract}
AM~CVn systems are ultra-compact, hydrogen-depleted and helium-rich, accreting binaries with degenerate or semi-degenerate donors. We report the discovery of five new eclipsing AM~CVn systems with orbital periods of 61.5, 55.5, 53.3, 37.4, and 35.4 minutes. These systems were discovered by searching for deep eclipses in the Zwicky Transient Facility (ZTF) lightcurves of white dwarfs selected using \gaia\ parallaxes. We obtained phase-resolved spectroscopy to confirm that all systems are AM~CVn binaries, and we obtained high-speed photometry to confirm the eclipse and characterize the systems. The spectra show double-peaked He-lines but also show metals, including K and Zn, elements that have never been detected in AM CVn systems before. By modelling the high-speed photometry, we measured the mass and radius of the donor star, potentially constraining the evolutionary channel that formed these AM~CVn systems. We determined that the average mass of the accreting white dwarf is $\approx0.8$\,\Msun, and that the white dwarfs in long-period systems are hotter than predicted by recently updated theoretical models. The donors have a high entropy and are a factor of $\approx$ 2 more massive compared to zero-entropy donors at the same orbital period. The large donor radius is most consistent with He-star progenitors, although the observed spectral features seem to contradict this. The discovery of 5 new eclipsing AM~CVn systems is consistent with the known observed AM~CVn space density and estimated ZTF recovery efficiency.

\end{abstract}




\section{Introduction}


AM~CVn systems are ultra-compact accreting binaries with degenerate or semi-degenerate, hydrogen-depleted and helium-rich donors. They are part of the family of cataclysmic variables; a primary white dwarf that is accreting mass from a donor via Roche lobe overflow \citep{warner1995}. For AM~CVn binaries, the donors are degenerate and the binary is very compact with orbital periods ranging from 65 to as short as 5 minutes (see \citet{solheim2010} for a review). Thousands of AM~CVn systems are expected to be present in our Galaxy, but their intrinsic faintness limits the known population to $\approx$ 60 AM~CVn systems (see \citealt{ramsay2018} for a recent compilation).

Because of their compactness and short orbital periods, AM~CVn stars are an excellent tool to study accretion physics under extreme conditions \citep[e.g.][]{kotko2012,coleman2018, cannizzo2019,oyang2021}.
Their short orbital periods also means that their angular momentum losses are dominated by gravitational wave radiation. Several hundred AM~CVn systems will be detectable by the Laser Interferometer Space Antenna \citep[LISA,][]{Amaro-Seoane2017} satellite and are one of the most abundant types of persistent LISA sources \citep{kremer2017,breivik2018,kupfer2018}. LISA will continuously observe these systems and will show the orbital period evolution of short-period AM~CVn systems.
AM~CVn systems are also potential progenitors of rare transient events. \citet{bildsten2006} discuss that, as a layer of He builds up on the accreting white dwarf, recurring He-shell flashes can occur which would look like He-Novae. 
The mass of the He-shell becomes larger and the time between flashes longer as the systems evolve to longer orbital periods. This can result in a very energetic `final-flash' which can be dynamical and eject material from the white dwarf, dubbed a `.Ia' transient \citep{shen2010}. The most important open question regarding AM~CVn systems is their formation channel, and how AM~CVn systems fit into the overall picture of compact binary evolution (see Section \ref{subsec:formationchannels} and also \citealt{nelemans2001,toloza2019}). 

In this paper, we present the search of Zwicky Transient Facility (ZTF) lightcurves to find new eclipsing white dwarf binaries and the discovery of five new AM~CVn systems and their characterization. In Section \ref{sec:background} we briefly summarize the current understanding of the formation channels and observational properties of AM~CVn systems, and discuss the currently known eclipsing systems.
Section \ref{sec:targetselection} presents the method and search of the ZTF lightcurves used to find the five new eclipsing AM~CVn systems. Section \ref{sec:observations} presents all the follow-up observations, high-speed photometry and phase-resolved spectroscopy, and archival data used to characterize the systems. Section \ref{sec:methods} presents the methods we used to characterize the systems, and Section \ref{sec:analysisresults} presents the results of this analysis. In Section \ref{sec:discussion} we discuss the implications of the measurements, specifically what the results imply for the formation channels of AM~CVn systems. We summarize this paper in Section \ref{sec:summary} and end with a short discussion on future work in Section \ref{sec:futurework}.

\section{Background}\label{sec:background}
\subsection{Formation channels}\label{subsec:formationchannels}
There are three proposed AM~CVn formation channels, but their relative importance is uncertain. In the white dwarf channel, a short period double white dwarf binary is formed after going through two common envelope phases \citep{ivanova2013}. The resulting binary is a typical CO white dwarf with a low mass helium white dwarf companion \citep{pac67,tutukov1989,deloye2007}. They evolve closer together as a result of gravitational wave radiation, and start stable mass transfer at orbital periods of $\approx$2-3\,m and evolve to longer orbital periods. Simulations by \citet{nelemans2001} predicted the white dwarf channel to be the dominant formation channel. However, \citet{shen2015} suggest that all double white dwarfs merge (because of friction with ejected material in a nova eruption), and double white dwarfs do not form AM~CVn systems. 

The second formation channel, the He-star channel, is similar to the white dwarf channel, but instead of a low mass white dwarf, the donor-progenitor is a non-degenerate He-burning star. These will also evolve to shorter periods, but start mass-transfer at slightly longer periods of 10 minutes \citep{savonije1986,iben1987,tutukov1989,yungelson2008,brooks2016}.

The third formation channel is the evolved-CV channel. In this scenario, the donor must evolve off the main sequence at the same time as the start of mass transfer. The system goes through a phase as a hydrogen-dominated CV (H-CV), and will eventually change into a helium-dominated CV once all the hydrogen has been stripped from the donor \citep[e.g.][]{podsiadlowski2003}. Models indicate that this channel is rare and does not contribute significantly to the AM~CVn population. This is due to the finely tuned starting parameters and long timescales required to remove all visible hydrogen from these systems \citep{nelson2018}.
However, there are known helium CVs which are potential progenitors of the channel \citep{ thorstensen2002,breedt2012,carter2013a,green2020}.

\subsection{Observational properties}
Although AM~CVn systems are all accreting DB white dwarfs with a degenerate or semi-degenerate donor, their observational characteristics vary significantly. Their appearance (both photometric and spectroscopic) is very different because of differences in the accretion rate, which is strongly correlated with the orbital period. The accretion rate determines the behaviour of the accretion disc \citep{kato2001,nelemans2004,ramsay2012,kupfer2013,coleman2018} and also sets the accreting white dwarf temperature \citep[e.g.][]{bildsten2006}. 

Very short period ($P\lesssim\!10$\,m) AM~CVn systems have high accretion rates and, depending on the masses, can be `direct impact' accretors. In these systems, there is no accretion disk and the accretion stream directly impacts the white dwarf. These systems emit X-rays, for example, HM Cnc and V407 Vul \citep{marsh2004}. 
Systems with periods $\gtrsim\!10$\,m form accretion discs. System with periods of $\approx\!10$--$22$\,m of the high accretion rate, the systems are in a constant `high state' \citep{kotko2012}. 
Intermediate period systems ($\approx\!22$--$45$\,m) with lower accretion rates show dwarf nova outbursts and feature large amounts of flickering in their lightcurves \citep[see][]{duffy2021}. As the orbital period increases, the outburst recurrence time increases exponentially \citep{levitan2015}, and the luminosity of the disk decreases as the orbital period decreases. 
In long-period systems ($\gtrsim\!45$\,m), the accretion rate is low, outbursts are very rare (recurrence times of $>\!100$ years), and the accreting white dwarf dominates the luminosity \citep{bildsten2006}.

\citet{ramsay2018} give an overview of the currently known sample of AM~CVn systems. This sample has been built up using various methods. Many AM~CVn systems (including AM~CVn itself) were identified by their blue colour and identification spectra. This method was used most recently by \citet{carter2013} who used SDSS colour information. The second main method of finding AM~CVn systems is by their outbursts. \citet{breedt2012,breedt2014} obtained spectra of Catalina Real-time Transient Survey (CRTS) dwarf novae and found three new AM~CVn systems. \citet{levitan2015} used the Palomar Transient Factory (PTF) to identify cataclysmic variables and identified AM~CVn systems with follow-up spectroscopy. \citet{vanroestel2021a} combined the colour selection and outburst selection methods and identified 9 new AM CVn systems.
\citet{isogai2019} also focused on outbursting CVs but instead used high-cadence photometry to find the period and identify a system as an AM~CVn system. Searching for short period variability was also used to identify a new AM~CVn system using Kepler data \citep{fontaine2011, kupfer2015, green2018a}.
Finally, in parallel with this work, \citet{burdge2020} searched for compact binaries by searching for short-period variability in Zwicky Transient Facility data. 

\subsection{Eclipsing AM~CVn systems}
Among the sample of known AM~CVn systems, only a few have been confirmed as eclipsing. Eclipsing AM~CVn systems where the primary white dwarf is eclipsed by the donor are extremely valuable since they allow the binary parameters to be measured. \citet{hardy2017} provides examples of modelling eclipsing accreting cataclysmic variable lightcurves with various accretion rates.

Three eclipsing AM~CVns have been studied in detail so far. The first is PTF1J1919+4815, in which just the edge of the disk and bright spot are eclipsed \citep{levitan2014}. The second is YZ~LMi where the white dwarf is partially eclipsed \citep{anderson2005,copperwheat2011}. Gaia14aae (ASASSN-14cn), is the only published AM~CVn in which the white dwarf is confirmed to be fully eclipsed by the donor \citep{Campbell2015,green2018,green2019}. Besides these well-studied cases, there are also a number of promising candidates. \citet{burdge2020} presented the discovery of a candidate 17.20 orbital period eclipsing AM~CVn system also using ZTF data. In addition, follow-up of ES Cet by \citet{bakowska2020} indicates that this system is also eclipsing, although it seems that in this case only the disk is eclipsed and not the accreting white dwarf.

Because for YZ LMi and Gaia14aae the white dwarf is eclipsed, the donor mass and radius have been measured to a few per cent accuracy by modelling the lightcurve. These precise measurements and their comparison with models show an interesting result \citep{copperwheat2011,green2018}. The donor for YZ LMi is relatively large and is consistent with either the helium star channel or the white dwarf channel. The donor in Gaia14aae is large, which is consistent only with mass-radius models describing the evolved CV channel. This is a surprising result since the evolved CV channel is expected to be the most uncommon channel of the three. These studies demonstrate that eclipsing AM~CVn systems allow us to precisely characterize them and better understand their evolutionary history. This motivated us to perform a dedicated search for eclipsing AM~CVn systems.

\section{Target selection}\label{sec:targetselection}

\begin{table*}
\caption{Overview of the five new eclipsing AM~CVn stars discovered by their eclipses using ZTF lightcurves. The parallax is taken from \gaia\ eDR3 \citet{brown2020}. The distance is calculated using a prior based on the white dwarf population \citep{kupfer2018}. The dust-extinction is taken from \citet{green2019a}. } \label{tab:overview}
\renewcommand{\arraystretch}{1.25}
\begin{tabular}{lllllllllll}
Name & RA & Dec & $P_\mathrm{orb}$ & ZTF-$g$  &  ZTF-$r$  & $G$ & $BP-RP$ & parallax & distance &  E(g-r) \\
 & & & minutes & AB-mag & AB-mag & Vega-mag & Vega-mag & mas & pc & \\
\hline\hline
\A & 16$^{\rm h}$37$^{\rm m}$43.6$^{\rm s}$ &  \phantom{+}49$^{\circ}$17$^{'}$40.9$^{''}$ & 61.5 & 19.34 & 19.49 & 19.40 & 0.27 & 4.88 $\pm$ 0.27 & $207^{+8}_{-8}$ & $<0.01$ \\
\E & 00$^{\rm h}$03$^{\rm m}$22.4$^{\rm s}$ &  \phantom{+}14$^{\circ}$04$^{'}$59.0$^{''}$ & 55.5 & 20.19 & 20.14  & 20.20 & 0.35 & 4.03 $\pm$ 0.54 & $263^{+29}_{-40}$ & $0.10^{+0.02}_{-0.02}$\\
\B &  02$^{\rm h}$20$^{\rm m}$08.6$^{\rm s}$ &  \phantom{+}21$^{\circ}$41$^{'}$55.8$^{''}$ & 53.3 & 19.72 & 19.82 & 19.76 & 0.09 & 2.94 $\pm$ 0.50 & $350^{+39}_{-58}$ & $0.14^{+0.02}_{-0.02}$\\
\C & 22$^{\rm h}$52$^{\rm m}$37.1$^{\rm s}$ &  \textminus{}05$^{\circ}$19$^{'}$17.4$^{''}$ & 37.4 & 19.05 & 19.16 & 19.09 & 0.04 & 1.95 $\pm$ 0.31 & $536^{+82}_{-93}$ & $0.01^{+0.01}_{-0.02}$\\
\D &  04$^{\rm h}$07$^{\rm m}$49.3$^{\rm s}$ &  \textminus{}00$^{\circ}$07$^{'}$16.7$^{''}$ & 35.4 & 19.23 & 19.41 & 19.42 & 0.07 & 1.33 $\pm$ 0.37 & $810^{+190}_{-230}$ & $0.14^{+0.01}_{-0.02}$\\
\hline
\end{tabular}
\end{table*}

\begin{figure*}
	\includegraphics{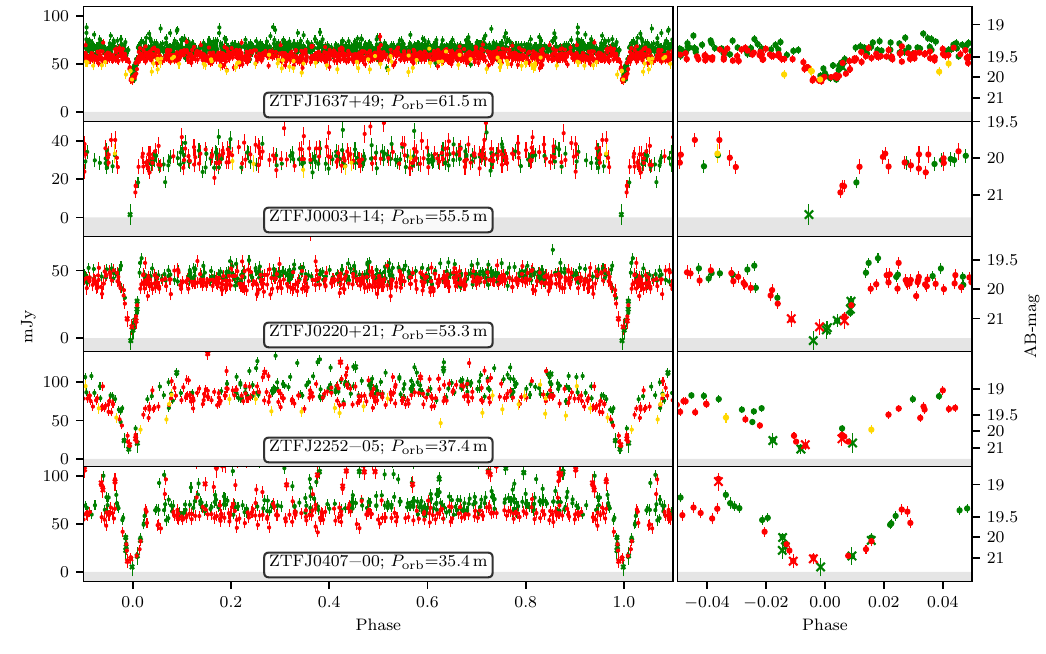}
    \caption{The ZTF lightcurves of the five new AM~CVn stars 
    folded on their orbital periods. ZTF $g$, $r$, and $i$-bands are displayed in green, red, and gold. Dots show ZTF PSF-photometry, crosses show ZTF alert photometry. The narrow eclipse in each of the lightcurves was used to discovery these objects. The two short period systems also show irregular variability in their lightcurves (see Section \ref{subsec:longtimescale} and Figure \ref{fig:longLC}).}
    \label{fig:ZTFlcs}
\end{figure*}

\subsection{Zwicky Transient Facility lightcurves}
As part of the Zwicky Transient Facility (ZTF), the Palomar 48-inch (P48) telescope images the sky every night \citep{graham2019,bellm2019,dekany2020}. Several sub-surveys are carried out by ZTF \citep{bellm2019b}, including an all-sky survey (publicly available), but also higher cadence surveys of smaller areas \citep[e.g.][]{vanroestel2019,kupfer2021}. Most surveys use $g$ and $r$ bands, but a small fraction of the observations are done in $i$ band. The exposure times are predominantly 30 seconds for $g$ and $r$ and 60 seconds for $i$ exposures. The median limiting magnitude, averaged over the lunar cycle, is \mbox{$r\approx20.5$} in all three bands. ZTF images are automatically processed and two main data products are generated. The first are the `alerts' which are based on difference imaging and are mainly designed to identify transients. The second main data product is PSF-photometry of persistent sources in the science images \citep[for a full description see][]{masci2019}.

\subsection{Search method}
To find deep eclipsing white dwarfs, we combined the PSF photometry data from the science images with the alert data generated from the difference images. We do this because the PSF-photometry is only reported for sources that are detected with a significance of 5-standard deviations in the science image. If the target shows a deep eclipse, no PSF-photometry is reported. Alerts are only generated for 5-sigma detection on the difference images, both positive or negative. A non-detection of a source thus generates a negative alert. The negative alerts (like the positive alerts) are vetted by a `real-bogus' system, which is based on deep-learning \citep{duev2019}.

To identify periodic eclipses we used the Boxed-Least-Square (BLS) algorithm\footnote{\url{https://github.com/johnh2o2/cuvarbase}} \citep{kovacs2002}. We normalised the $g$, $r$, and $i$, data and combined them into a single lightcurve, rejecting any points which were flagged as potentially bad ($5-10$\%, mostly due to clouds). In addition, we removed any points which were more than 5 standard deviations above the median level of lightcurve in order to reject outbursts. We searched a linear frequency grid and used a frequency step inversely proportional to the baseline and the minimum eclipse duration, and use an additional oversampling factor of 10. This typically results in a grid of a few million frequencies to be searched. To identify the best period, we simply used the period with the highest power. In the case of multiple best periods, we used the longest period.

As a post-processing step, we fit the phase-folded lightcurve with a simple trapezoid plus two sinusoids to model an eclipse, reflection effect, and ellipsoidal modulation. The eclipse width is shared between colour bands, but the depth of the eclipse and amplitude of any sinusoidal component is independently determined for each band. Periods, the period power, the goodness-of-fit of the model, and other statistics were saved for later selection of lightcurves for visual inspection.

\subsection{Application to the ZTF lightcurves}
We analysed the ZTF-lightcurves DR3 and private data (available on 2019 December 11th) of all the sources in the \gaia\ DR2 white dwarf catalogue from \citet{gentilefusillo2019}, in our search for eclipsing white dwarfs.
We processed all lightcurves with more than 80 epochs in $g$ and $r$ combined (241\,775 objects), and searched a period range from 0.02\,d (28.8\,m) to 5\,d on a linear scale in frequency. The lower limit was chosen to reduce the computational time and because \citet{burdge2020} has already searched for shorter period systems. We visually inspected the folded lightcurves for all objects in the white dwarf catalogue which had at least one alert associated with it (6\,412 objects). For the rest, we only inspected lightcurves for which the model fit showed an improvement of $\Delta \chi^2>50$ compared to fitting a straight line (27\,116 objects). In this work, we present the AM CVn systems found using this method, a full catalog of eclipsing white dwarf and other interesting discoveries \citep[e.g.][]{vanroestel2021b} will be presented in  future work.

Figure \ref{fig:ZTFlcs} shows the folded ZTF lightcurves of the five eclipsing AM~CVn systems we discovered in the sample of eclipsing white dwarfs. \A\ (also discovered by \citealt{keller2021}) is located in a region of the sky that is observed at a higher cadence by ZTF as part of a ZTF-partnership survey. For this reason, many more epochs are available compared to the rest of the sky. This allowed us to discover this object first, despite the eclipse being the most shallow and narrow of the five objects, and the fact that no alerts were generated for this object. \B\ and \D\ both show negative `alerts' in both $g$ and $r$ band. A visual inspection of their lightcurves confirmed their orbital period. \D\ also showed one outburst, which combined with its period, immediately confirmed its nature as an AM CVn system. \added{We note that \D\ was also identified as an outbursting cataclysmic variable by ASASSN and Gaia as ASASSN-18dg and Gaia20aby.}

\C\ was identified later in a general search of persistent sources which have a \gaia\ parallax \citep{bailer-jones2018} and are located close to the white dwarf track, and have negative ZTF-alerts. This sample contained a total of 11\,766 sources ZTF sources. 
\C\ triggered three negative alerts, and a BLS-period search of the lightcurve identified a potentially short period, which was confirmed using CRTS and PTF data. \C\ was not found in the initial search because its \gaia\ parallax was very uncertain, and not included in the catalogue by \citet{gentilefusillo2019}.

\E\ was identified as a candidate in the initial search but was initially not followed up because of the sparse ZTF-lightcurve and red $BP-RP$ colour. With more epochs available, the eclipse became more prominent and follow-up data confirmed its nature as an AM CVn system.

\section{Observations}\label{sec:observations}
For each system, we obtained high-speed photometry and phase-resolved spectroscopy. Table\,\ref{tab:observ} presents an overview of all follow-up observations. In addition, we used multiple surveys to study the long timescale evolution and spectral energy distribution.

\subsection{LRIS spectroscopy}
Phase-resolved spectroscopy of all systems was obtained using the Keck\,I Telescope (HI, USA) and the Low Resolution Imaging Spectrometer (LRIS; \citealt{Oke1995,McCarthy1998}).
We used a $1^{\prime\prime}$ slit, with the R600 grism for the blue arm, and R600 grating for the red arm. With this setup, spectral resolution is approximately $R\approx1100$ in the blue arm and $R\approx1400$ in the red arm. For one spectrum we used the R300 grating, which has a resolution of $R\approx800$ .

A standard long-slit data reduction procedure was performed with the \textsc{Lpipe} pipeline\footnote{http://www.astro.caltech.edu/\texttildelow dperley/programs/lpipe.html} \citep{Perley2019}. The pipeline reduced LRIS spectral data to spectra using the standard procedure, including calibration with a standard star. Wavelength calibration was done using lamp-spectra obtained at the beginning and end of each sequence. Fine-tuning of the wavelength calibration was done using sky-lines.

 In order to identify spectral lines, we interpolated the spectra to a common wavelength grid, averaged them and normalised the result using a 7th order polynomial fitted by minimizing the absolute error. The absolute error is less sensitive to emission and absorption lines compared to the squared error commonly used.


\subsection{CHIMERA fast cadence photometry}
CHIMERA \citep{harding2016} is a dual-channel photometer that uses frame-transfer, electron-multiplying CCDs mounted on the Hale 200-inch (5.1 m) Telescope at Palomar Observatory (CA, USA). The pixelscale is 0.28 arcsec/pixel (unbinned). We used the conventional amplifier and used 2x2 binning on most nights (except when the seeing was excellent) to reduce the readout noise and readout time. Each of the images were bias subtracted and divided by twilight flat fields\footnote{\url{https://github.com/caltech-chimera/PyChimera}}.

We used the ULTRACAM  pipeline to do aperture photometry \citep{dhillon2007}. We used an optimal extraction method with a variable aperture of 1.5 times the FWHM of the seeing (as measured from the reference star). A differential lightcurve was created by simply dividing the counts of the target by the counts from the reference star. Timestamps of the images were determined using a GPS receiver.

We obtained CHIMERA data for each target in the $g$ and $r$ or $i$ filters during dark or grey time (see Table \ref{tab:observ}). The seeing for the CHIMERA observations was typically between 0.8 and 1.5 arcsec.

\subsection{HiPERCAM fast cadence photometry}
HiPERCAM is a high-speed camera for the study of rapid variability \citep{Dhillon2021} and is able to obtain images in $ugriz$ simultaneously. HiPERCAM was mounted on the 10.4m Gran Telescopio Canarias on La Palma (Spain). With this telescope, HiPERCAM has a $0.081^{\prime\prime}$/pixel scale and a $3.1^\prime$ field-of-view across the diagonal. 

We obtained 76 minutes of data in $ugriz$ of \A. The observation was timed such that the lightcurves covers two eclipses. The data were reduced using the dedicated HiPERCAM pipeline\footnote{https://github.com/HiPERCAM/}, including debiasing and flat-fielding. 
Differential lightcurves were constructed using a single reference star. Timestamps were obtained from a GPS receiver.

\subsection{'Alopeke fast cadence photometry}
'Alopeke is a low noise, dual-channel imager mounted on the Differential Speckle Survey Instrument \citep[DSSI,][]{scott2019} of the 8m Gemini North Telescope on Mauna Kea (Hawaii, USA). In the blue arm we used the $g$ filter and in the red arm we used the $i$ filter. 
The pixelscale of the wide-field mode is 0.07 arcsec/pixel.

Alopeke was used to obtain 5 lightcurves of the eclipse of \B. Images were bias subtracted and flatfielded using the standard procedure. We used aperture photometry to extract lightcurves using the ULTRACAM pipeline. No bright stars were close enough to use differential photometry. Instead, we used a large aperture of 2.5 times the average FWHM, and used the raw lightcurve. No GPS receiver was available for Alopeke, so we used computer timestamps instead.

\subsection{Archival photometry}
To study the long timescale photometric variability and the spectral energy distribution, 
we obtained single epoch or averaged photometry data from multiple other survey telescopes: \textit{Galex} \citep{GALEX2017}, \gaia\ eDR3 \citep{GaiaeDR3}, and \textit{WISE} \citep{CatWISE2020}. We also obtained multi-epoch photometry from SDSS \citep{SDSSDR13}, Pan-STARRS \citep{PS1}, CRTS \citep{drake2009}, ASAS-SN \citep{shappee2014,kochanek2017}, PTF \citep{law2009,rau2009} and ATLAS \citep{tonry2018,smith2020}.

\section{Methods}\label{sec:methods}

\subsection{Timing}\label{subsec:timing}

To determine the ephemerii of the mid-eclipse times, we use both the ZTF lightcurve, archival data (SDSS, PS, and CRTS) and the follow-up photometry.
We fit the normalised and combined ZTF $g$, $r$, and $i$ lightcurves with the best \textit{lcurve} model (see Section \ref{sec:lcfit}) as determined from the high-cadence photometry. The only free parameters are the period and mid-eclipse time. As a prior, we use the measured mid-eclipse times from the CHIMERA, HiPERCAM, and Alopeke lightcurves. In addition, we also add the epochs of any individual in-eclipse points from SDSS, Pan-STARRS and PTF, and use half their exposure time as the uncertainty.
The uncertainties (standard deviations) are determined by using \textit{emcee} \citep{ForemanMackey2013}. 

\subsection{Doppler tomography}
To better understand the geometry of the system, we use Doppler tomography \citep{marsh2001}. Doppler tomography converts a series of 2D spectra obtained at different phases of the orbit into a velocity map (also called a Doppler map). We used the He-\textsc{I} lines at 6678.15, 7065.17, and 7281.35\,\AA\ for \A, \B, and \D, 
 because these are not blended with other lines. For \E\ and \C, more blue spectra are available and in this case, we use He-\textsc{I} lines in the blue spectra: 3613.64, 3634.23, 3888.64, 4026.19, and 4921.93\AA.
 To make the Doppler maps from the LRIS spectra, we used the python package \textsc{doppler}\footnote{\url{https://github.com/trmrsh/trm-doppler}}. We used a blurring scale of 40\,$\mathrm{km s^{-1}}$, and manually set the target reduced $\chi^2$ value to avoid over- or underfitting the map.

\subsection{Modelling of the spectral energy distribution}\label{subsec:SEDfit}
To estimate the white dwarf temperature, we fit the spectral energy distribution of the systems by simply modelling them as a blackbody. The model parameters are the temperature ($T_\mathrm{eff}$), the radius of the white dwarf ($R_\mathrm{WD}$), the distance ($d$), and the amount of reddening using the \citet{fitzpatrick1999} reddening law with the parameter $E_{B-V}$.

We constrain the distance by putting a Gaussian prior directly on the parallax using \gaia\ eDR3 values. We also use a Gaussian prior on the radius of the white dwarf using the values obtained using the lightcurve modelling. Finally, we constrain the value of $E_{B-V}$ using the Pan-STARRS dustmap \citep[Bayestar19,][]{green2019a}, with the conversion $E_{B-V} = 0.884 E_{g-r}$. 

We converted the magnitudes to flux using the zeropoints for each filter\footnote{\url{http://svo2.cab.inta-csic.es/theory/fps/}} \citep{rodrigo2012,rodrigo2020}. We use the effective wavelength for each filter to calculate the model flux. 

\begin{figure*}
	\includegraphics[width=\textwidth]{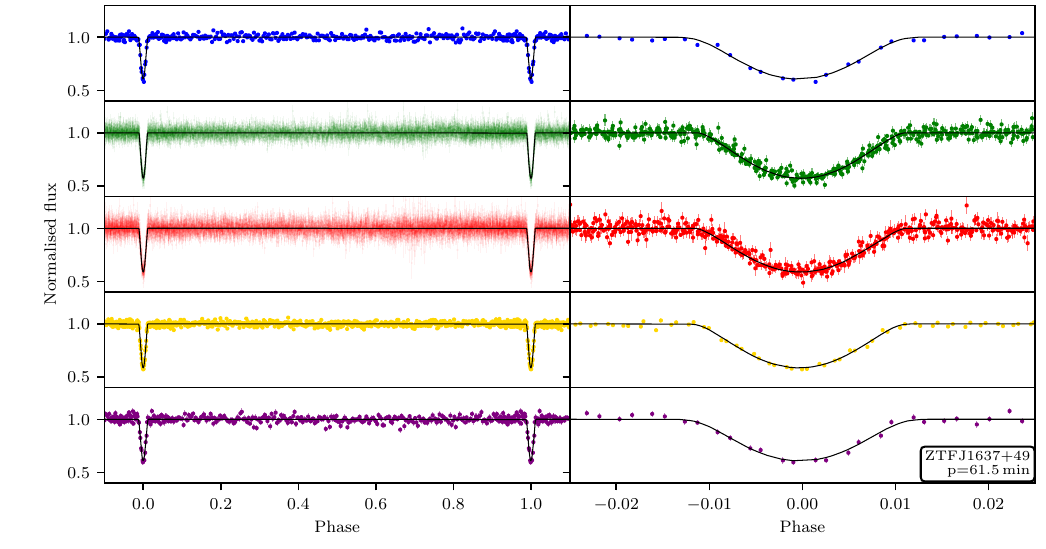}
    \caption{The lightcurves of \A\ with the best-fit models overplotted in the $ugriz$ bands (blue, green, red, gold, and purple; from top to bottom). The eclipse depth is identical is each of the 5 bands, which suggests that there is no contribution from the disk or brightspot. Therefore, the lightcurve model (black line) does not include these components, and the model only includes a white dwarf that is eclipsed by a cold and dark, Roche-lobe filling donor.}
    \label{fig:lcfit_A}
    \vspace*{\floatsep}
	\includegraphics[width=\textwidth]{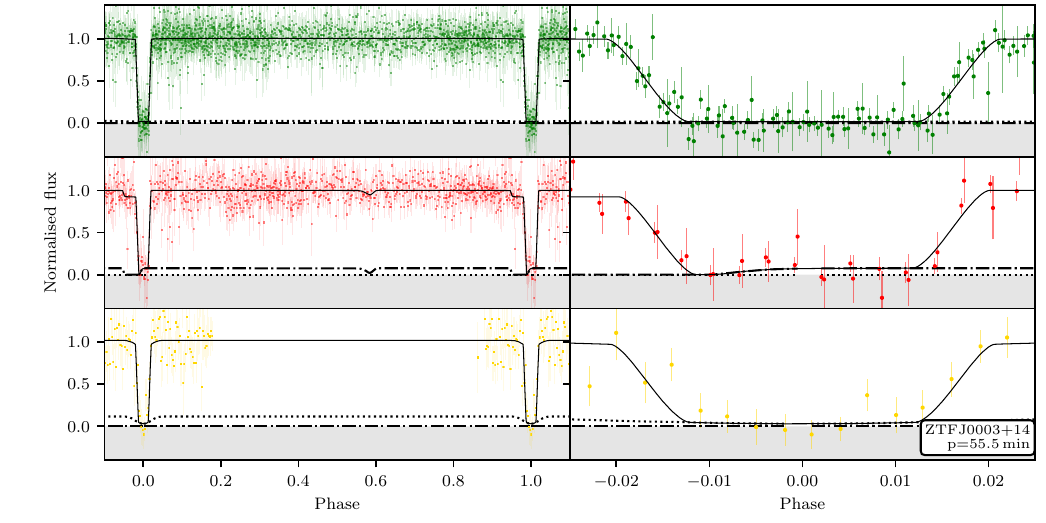}
    \caption{The combined lightcurves of \E\ in the $gri$ bands, similar to \ref{fig:lcfit_A}. The full model is shown by the black line, the contribution from the disk is shown by the dotted line, and the contribution by the brightspot is shown by the dashed-dotted line. The model shows that the white dwarf dominates the luminosity, and disk and/or brightspot only contribute $\lesssim 10\%$ to each lightcurve (see also Table \ref{tab:lumfrac}).}
    \label{fig:lcfit_E} 
\end{figure*}

\begin{figure*}
	\includegraphics[width=\textwidth]{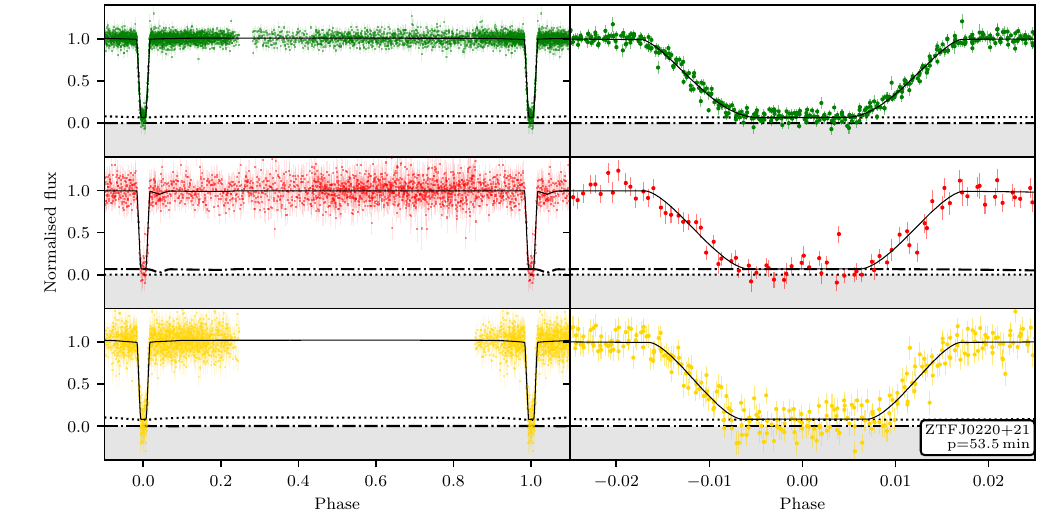}
    \caption{The combined lightcurves of \B\ in $gri$-bands (top to bottom, see also Figure \ref{fig:lcfit_E}). Similar to \E, the white dwarf dominates the luminosity and disk and/or brightspot only contribute $\lesssim 10\%$.}
    \label{fig:lcfit_B}
        \vspace*{\floatsep}
	\includegraphics[width=\textwidth]{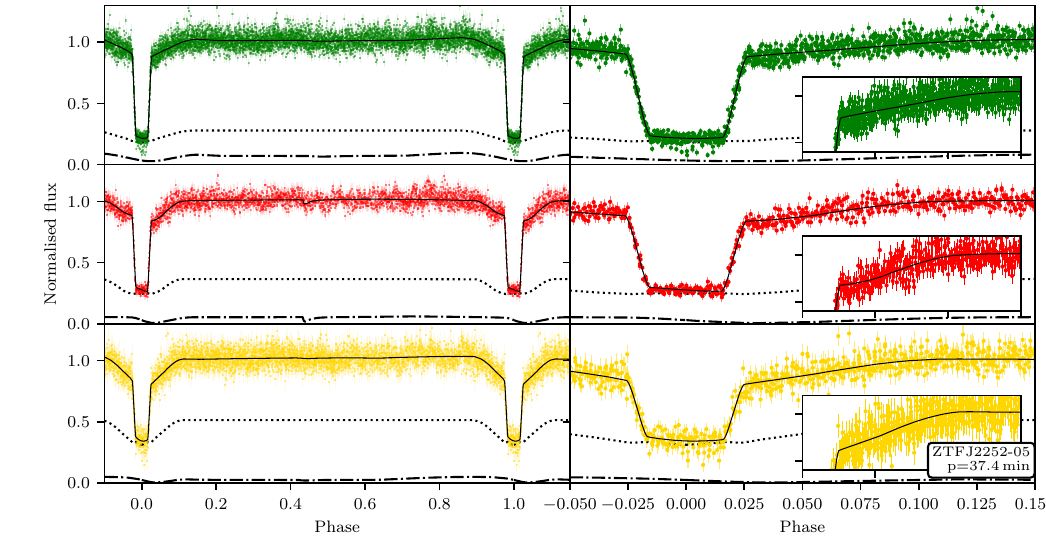}
    \caption{The combined lightcurves for \C in the $gri$-bands (top to bottom). The full model is indicated by the black line, while the contribution from the disk and brightspot are indicated with a dotted and dashed-dotted line. The lightcurves have been detrended, but flickering remains visible. The disk contributes significantly more compared to the longer period systems (27--51\%), with larger contributions for longer wavelengths. In each of the lightcurves, a few percent (3--7\%) contribution from the brightspot is also visible, shown in the inset.}
    \label{fig:lcfit_C}
\end{figure*}

\begin{figure*}
	\includegraphics[width=\textwidth]{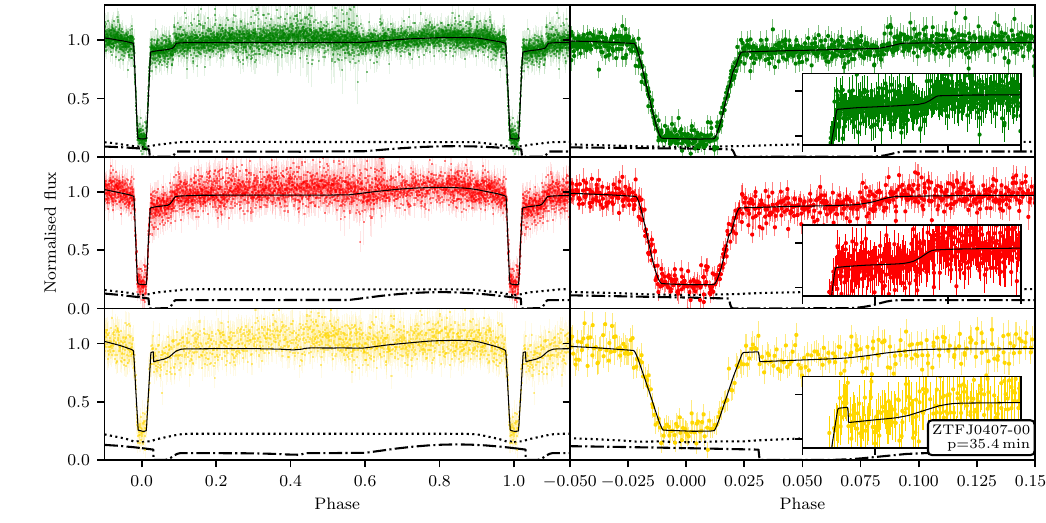}
    \caption{The combined lightcurves of \D\ in $gri$-bands (top to bottom, similar to Figure \ref{fig:lcfit_E}). Compared to \C\, the disk contributes less for this system (13--22\%). The overall contribution by the brightspot is similar as for \C\, but the eclipse is clearly visible in the $g$ and $r$ lightcurves.}
    \label{fig:lcfit_D}
\end{figure*}

The observed flux is the combined flux from the white dwarf, disk, brightspot, and the donor. Using the $gri$ lightcurves of the fully eclipsing systems, we can disentangle the contributions from the different components by modelling the lightcurves in those three bands (see Figures \ref{fig:lcfit_A}--\ref{fig:lcfit_D}). For each system, we use the model to determine how much the disk and brightspot contribute. We use this to correct the $gri$ magnitudes from PS, SDSS, and ZTF magnitudes. 

To find the optimal solution and uncertainties we use \textit{emcee}. We use a likelihood function assuming Gaussian uncertainties. In order to account for any systematic errors, we include an additional error term in the model which is optimized together with the other parameters. The measurements in other bands are used as upper limits (implemented as one-sided Gaussians). We exclude the two \textit{WISE} bands from the fit as these are likely dominated by the disk and/or donor. 

For \A\ and \E\ we deviated from this general approach. 
\A\ only shows partial eclipses, and we cannot use the eclipse depth and shape to disentangle the contribution to the overall luminosity by the different components. We use the uncorrected magnitudes to fit the white dwarf model and assume the other components can be neglected, which is justified because this is a very long period system with a very low accretion rate and cold donor. We exclude the SDSS $r$ band measurement which was obtained during an eclipse and is 0.5\,mag fainter than other $r$-band observations.

For \E, the SDSS data was significantly brighter than the PS and \gaia\ data, which suggests that the system was in outburst when SDSS observed it. We, therefore, use the SDSS $ugriz$ data as upper limits only.

\subsection{Lightcurve modelling} 
\label{sec:lcfit}
We use \textit{Lcurve}\footnote{\url{https://github.com/trmrsh/cpp-lcurve}} to model the lightcurves and infer the binary parameters. \textit{Lcurve} uses 3D grids to simulate two stars (spherical or with a Roche geometry). In addition, a disk and bright-spot can be added to the model. In this section, we briefly summarize the model setup. For a detailed description of \textit{Lcurve} is given by \citet{copperwheat2011}, and a discussion on the subtleties of modelling eclipsing AM CVn lightcurves, we refer the reader to \citet{green2018}.

 The basic model is a spherical white dwarf with a dark, Roche-lobe filling donor. The free parameters for this model are the inclination ($i$), mass-ratio ($q$, which sets the radius of the donor, $r_2$), scaled radius of the accreting white dwarf ($r_1$), the mid-eclipse time ($t_0$), and the velocity scale ($[K_1+K_2]/\sin i$), with $K_{1,2}$ the observable radial velocity amplitude. We fixed the temperature of the white dwarf to the values obtained from the fit to the SED (Section  \ref{subsec:WDtemp}). We set the donor temperature to $2000$\,K, which in practice means that the donor does not contribute to the model lightcurve. 
We fixed the orbital period to the period derived from the ZTF data (see Section \ref{subsec:timing}).
Limb-darkening of the white dwarf is approximated by using the 4-parameter Claret law \citep{claret2011} with the parameters taken from the DB models by \citet{claret2020} for the closest temperature value and $\log(g)=8.5$. In addition, we imposed two restrictions on the white dwarf radius using the approximation of the mass--radius relation of Eggleton from \citet{rappaport1989}. The first is that it cannot be smaller than a zero-temperature white dwarf. 
The second is a Gaussian prior on the white dwarf radius relative to the white dwarf M-R relation with an uncertainty of 5\%.

The lightcurves show a range in contributions from the disk and brightspot, and therefore we use different model configurations and free parameters for each of the AM CVn systems.  At very long orbital periods, both the disk and bright spot components do not contribute significantly to the lightcurve, as is the case for \A. This is supported by the fact that the eclipse depths are identical in each of the 5 HiPERCAM bands. We therefore use just the basic model without a disk or brightspot to model the lightcurve. For \E\ and \B\, we include both a disk and brightspot in the model to account for any small amount of in-eclipse light. We use the same setup as in \citet{green2018} and keep most parameters fixed. The only two free parameters of the disk are the temperature ($T_\mathrm{disc}$) and the outer radius ($r_\mathrm{disc}$). For the brightspot, we keep most parameters fixed to their default values. The only two free parameters of the brightspot are the temperature ($T_\mathrm{spot}$) and spot length ($l_\mathrm{spot}$). We fix the spot position to the outer radius of the disk. For \C\ and \D\, we used a model with a disk with two free parameters ($T_\mathrm{disc}$, $r_\mathrm{disc}$), and use five free parameters for the spot; length, exponent, angle, yaw, temperature, and cfrac (again, see \citealt{green2018} for a description of each parameter). We again fixed the brightspot location to the outer edge of the disk.

To find the best parameter values and uncertainties, we use a Markov chain Monte Carlo (MCMC) method as implemented in \textit{emcee}. Before fitting the data, we removed long-timescale trend by fitting a polynomial to the individual lightcurves. For the two short-period systems, we removed any remaining flickering using Gaussian process regression, using a 3/2 Matern Kernel. Finally, we rescaled the uncertainties in the lightcurves so they account for any remaining difference between the data and model. 
For each lightcurve, we use 512 walkers and at least 2000 generations to determine the best-fit model and the uncertainties on the parameters.

\section{Results}\label{sec:analysisresults}

\subsection{Ephemerii}
The mid-eclipse ephemerii are given in Table \ref{tab:ephemerii}. The combination of archival data that spans a long-baseline and accurate eclipse time measurements means that the periods can be constrained to $\approx 10^{-8}$ days or better, and the zeropoint to a few seconds. We did not detect any significant deviations from the eclipse arrival time, except for system for \D. More data is needed to verify these eclipse arrival time variations and determine their nature, which we will be the topic of future work.

\begin{table}
    \centering
    \begin{tabular}{lll}
     Name & $t_0$ ($\mathrm{BJD_{TDB}}$) & period (d) \\
    \hline\hline
    \A & 2458370.73498(4) & 0.042\,707\,771(8) \\
    \E & 2458323.02453(1) & 0.038\,564\,45(8) \\
    \B & 2458428.74931(8) & 0.037\,041\,443(2) \\
    \C & 2458469.64845(7) & 0.025\,968\,133\,6(9) \\
    \D & 2458386.99699(2) & 0.024\,588\,123(3) \\
    \hline
    \end{tabular}
    \caption{The orbital period and mid-eclipse times as measured from the ZTF and archival photometry.}
    \label{tab:ephemerii}
\end{table}


\subsection{Dopplermaps}

\begin{figure*}
\centering
\includegraphics[scale=0.99]{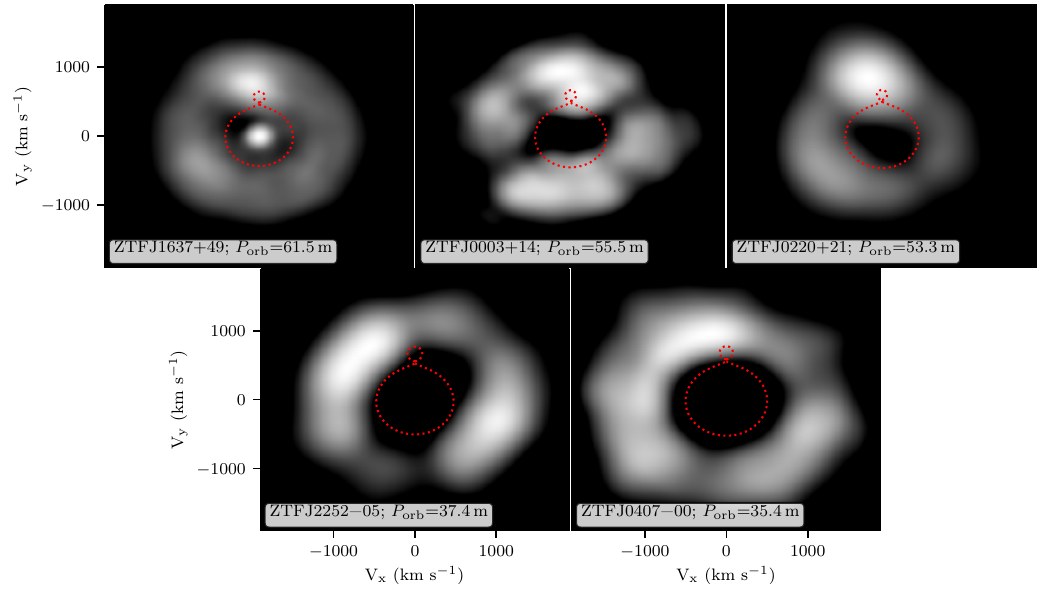}
\caption{Dopplermaps of the five eclipsing AM CVn systems. For \A, \B, and \D\ we used spectra from the LRIS red arm, and for \E\ and \C\ we used spectra from the LRIS blue-arm. Red dashed lines show the Roche lobes of the binary system, with the white dwarf located in the center of the image and the donor (with the small Roche-lobe) on top. In this coordinate system, the accretion flow originates at the top of the donor. \A shows emission from the white dwarf (the 'central-spike'). The primary brightspots are visible at the top or top-left of the figure for all systems. \A\ and \B show possible secondary spots at the bottom-left, and \C\ shows a possible secondary spot at the bottom right. The structure in the map of \E\ is caused by the sparse sampling. }
\label{fig:dopplermaps}
\end{figure*}

The Dopplermaps are shown in Figure \ref{fig:dopplermaps}. All systems show a clear disk in the Dopplermaps. The size of the disk increases slightly as the orbital decreases and the velocities of the components increase.
For \A\, and \B, (with the best data) show a brightspot at the top of the figure, and possibly a secondary brightspot at the lower right of the map, also seen in Gaia14aae \citep{green2020}. `Central spike' emission is clearly visible for \A\ which originates from the white dwarf \citep{kupfer2016,green2020}. The other Dopplermaps are less clear, but all seem to show brighter parts in the disk that are consistent with bright-spots.

\subsection{Spectral energy distribution and white dwarf temperature}\label{subsec:WDtemp}

\begin{figure*}
    \centering
    \includegraphics{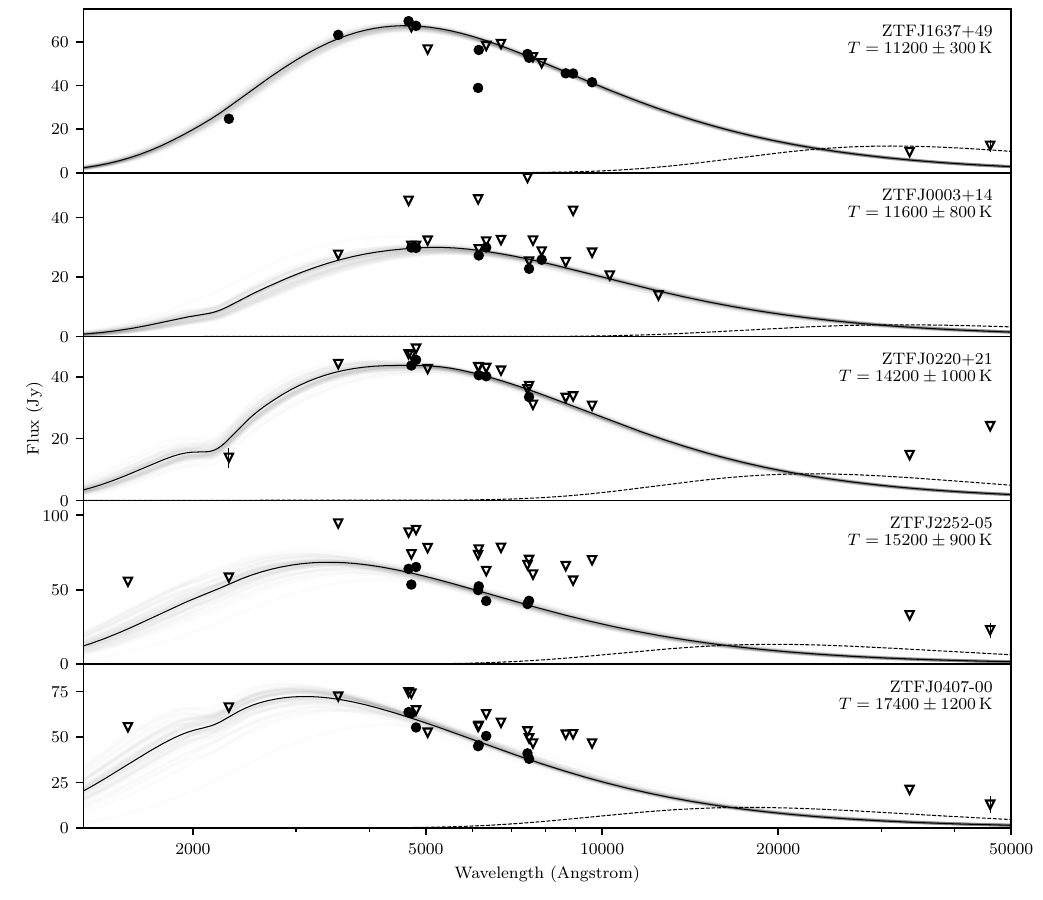}
    \caption{The SED of each system, showing \textit{Galex}, SDSS, PS, median ZTF, UKIDSS, and \textit{WISE} data. Triangles show \textit{observed} flux measurements which were used as upperlimits in the fit. Filled dots show the estimated flux by the white dwarf by removing the contributions from the disk and brightspot (Table \ref{tab:lumfrac}). Errorbars are omitted if they are smaller than the symbol size. We fitted the data with a blackbody in order to measure the white dwarf temperature (black line). Grey lines show the uncertainty in the fit. The dashed line shows the expected contribution from the donor star if we assume an equilibrium temperature that is set by irradiation from the white dwarf. For \A\ the correction of the observed flux is assumed to be 0.}
    \label{fig:SED}
\end{figure*}

The spectral energy distributions are shown in Figure \ref{fig:SED}. Also shown are the best-fit white dwarf models (see Section \ref{subsec:SEDfit}), and the white dwarf temperatures as reported in Table \ref{tab:allbinarystats}. In general, the models agree well with the data, but do show some inconsistencies at extreme wavelengths. For shortest period systems, a FUV measurement is available and the model underestimates the FUV flux in both cases. This is possibly due to emission from the accretion disk, but could also be due to the blackbody assumption. 

The flux in the \textit{WISE} bands, which were not used to fit the model, are also underestimated in all cases. To check if this would be consistent with emission from the donor, we calculated the donor equilibrium temperature based on irradiation by the white dwarf. This suggests that the excess in the far IR for \A\ could be by the donor (see also \citealt{green2020}). For the other objects, the estimated contribution by the donor is insufficient to explain the excess light, and the disc is the more likely source of the IR excess.

\subsection{Long timescale variability}\label{subsec:longtimescale}
\begin{figure*}
\includegraphics[width=\textwidth]{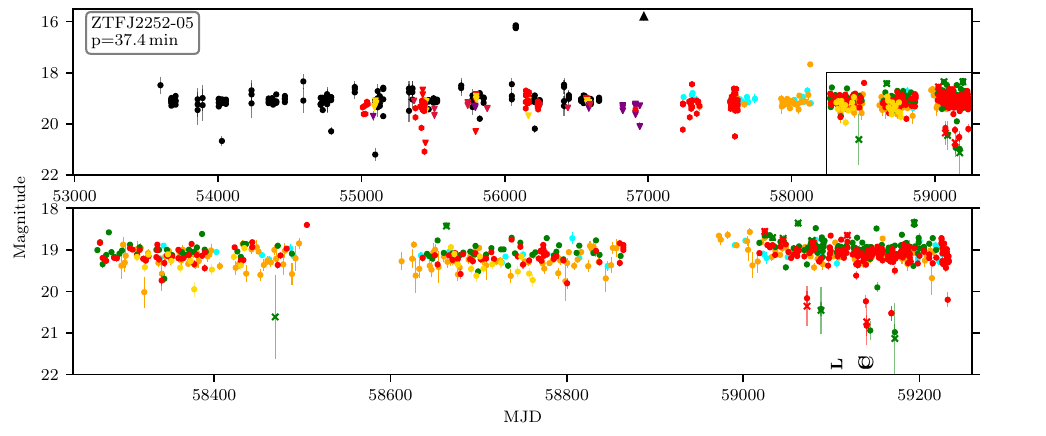}
\includegraphics[width=\textwidth]{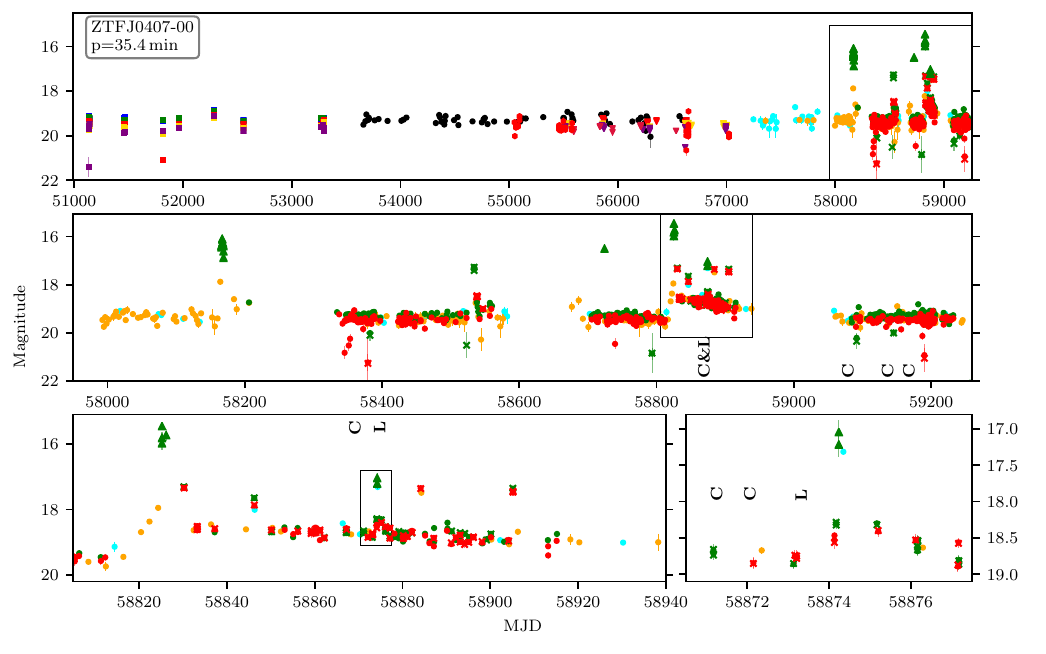}
\vspace{-0.5cm}
\caption{The lightcurves of the two shortest period systems \C\ (top) and \D\ (bottom). The consecutive panels show an increasingly zoomed in range, indicated by black boxes in the previous panels.
The figures show photometry from SDSS (squares), CRTS (dots), PTF (hexagons), ASAS-SN (triangles) and ZTF data (PSF photometry as dots and alerts as crosses). The $ugriz$ filters are shown in blue, \replaced{green, red,}{red, green,} yellow and purple, $V$-band observations are shown in black. "C" and "L" indicate when Chimera and LRIS data were obtained. 
\C: Two outbursts were detected by CRTS and ASAS-SN, three years apart. No other outbursts can be seen in the rest of the data, including well-sampled ATLAS and ZTF data of the last 5 years.
\D: No outbursts or long-timescale variability can be identified in the SDSS, CRTS and PTF data. The ATLAS cyan and orange filter data are shown in cyan and orange dots, with the data binned into one-day bins. The middle panel shows ZTF and ASAS-SN data from 2017 to 2020 which shows two large-amplitude, multi-day long outbursts. The bottom panel shows the most recent data, which shows 4 short-duration ($\lesssim$1 day) outbursts which occur very shortly after a superoutburst. These echo outbursts have also been seen in other AM CVn systems \citep[e.g.][]{green2020,duffy2021,pichardomarcano2021}.}
\label{fig:longLC}
\end{figure*}

We inspected the ZTF and archival data to study the long-timescale behaviour of the AM CVn systems. As expected, the long period systems, \A, \E, and \B, do not show any outburst or other long timescale variability in the ZTF data or the other archival data that spans a decade or more. For \E, only ZTF data are available that spans three years, and no outbursts or trends are visible. We do note that the source is $\approx$0.6 mag brighter in SDSS compared to ZTF and Pan-STARRS data. \E\ likely showed an outburst, and SDSS observed it while it was on its way back to quiescence.

Both \C\ and \D\ do show variability on longer timescales, shown in Figure  \ref{fig:longLC}.
\C\ shows low-level variability in the lightcurve (besides the eclipse), but no clear overall trends. Only two outbursts have been detected, one by CRTS and one by ASAS-SN, both reaching magnitude 16. The lightcurves are not sampled well enough to make any statement on the duration of the outbursts. 

\D\ also shows variability on longer timescales. No outbursts have been detected by SDSS, CRTS, Pan-STARRS, or PTF.
The ZTF and ASAS-SN data obtained in the last three years show two (possibly three) high amplitude outbursts that last multiple days. As can be seen in the middle panel of Figure  \ref{fig:longLC} for \D, the quiescence luminosity increased after the last superoutburst by about 0.6 magnitudes and is slowly decreasing. This system shows at least 4 short duration (<1 day) outbursts.


\subsection{Binary parameters}

\begin{table*}
    \centering
    \begin{tabular}{l|ccccccccc}
    & $P$ &  $q$ & $i$ & $T_1$& $M_1$ & $M_2$ & $R_1$ & $R_2$ \\ 
    & min &   & $\deg$ & kK & \Msun & \Msun & \Rsun & \Rsun \\ 
    \hline
    \hline
    \A & 61.5 & $0.026\pm0.010$ & $82.7\pm0.09$ & $11.2\pm0.3$ & $0.90\pm0.05$ & $0.023\pm0.008$ & $0.009\pm0.001$ & $0.068\pm0.007$ \\ 
    \E & 55.5 & $0.0214\pm0.010$ & $86.5\pm2.0$ & $11.6\pm0.8$ & $0.79\pm0.11$ & $0.017\pm0.011$ & $0.0106\pm0.001$ & $0.057\pm 0.012$ \\ 
    \B & 53.5 & $0.0174\pm0.005$ & $85.3\pm2.1$ & $14.2\pm1.0$ & $0.83\pm0.07$ & $0.014\pm0.006$ & $0.010\pm0.001$ & $0.054\pm0.007$ \\ 
    Gaia14aae & 49.7 & $0.0290\pm0.0006$ & $86.27\pm0.10$  & $17.0\pm1.0$ & $0.872\pm0.007$ & $0.0253\pm0.0007$ & $0.00924 \pm 0.00009$ & $0.0603\pm0.0003$ \\ 
    \C & 37.4 &  $0.034\pm0.006$ & $87.0\pm1.0$ & $15.2\pm0.9$ & $0.76\pm0.05$ & $0.026\pm0.008$ & $0.010\pm0.001$ & $0.049\pm0.004$ \\ 
    \D & 35.4 & $0.024\pm0.004$ & $86.5\pm0.7$ & $17.4\pm1.2$ & $0.79\pm0.06$ & $0.019\pm0.003$ & $0.010\pm0.001$ & $0.044\pm0.002$ \\ 
    YZ LMi & 28.3 & $0.041\pm0.002$ & $82.6\pm0.3$ & $17.0\pm1.0$ & $0.85\pm0.04$ &  $0.035\pm0.003$ & - & $0.047\pm0.001$  \\ 
    \hline
\end{tabular}
    \caption{Binary parameters of all well-characterized eclipsing AM CVn systems. The 5 ZTF systems are from this work, parameters of Gaia14aae are from \citet{green2018}, and parameters of YZ LMi are from \citet{copperwheat2011}. In each case, we have used the results based on modelling of the $g$-band lightcurve, which is the most accurate in all cases.}
    \label{tab:allbinarystats}
\end{table*}

\begin{table*}
    \centering
    \begin{tabular}{cccccccccc}
    & \multicolumn{3}{c}{WD} & \multicolumn{3}{c}{disc} &
    \multicolumn{3}{c}{brightspot} \\
    & $g$ & $r$ & $i$ & $g$ & $r$ & $i$ & $g$ & $r$ & $i$ \\ 
    \hline
    \hline
        \A & 100\% & 100\% & 100\% & - & - & - & - & - & -  \\
        \E & 97\% & 93\% & 89\% & 3\% & 0\% & 11\% & 0\% & 7\% & 0\%  \\
        \B & 92\% & 93\% & 89\% & 8\% & 0\% & 10\% & 0\% & 7\% & 0\%  \\
        \C & 65\% & 58\% & 46\% & 27\% & 36\% & 51\% & 7\% & 5\% & 3\%  \\
        \D & 82\% & 75\% & 70\% & 13\% & 16\% & 22\% & 5\% & 9\% & 8\%  \\
    \hline
    \end{tabular}
    \caption{The contribution to the overall luminosity of white dwarf, disk, and brightspot in the $gri$ filter for the best-fitting models. The white dwarf is typically dominant for long period systems and at shorter wavelengths, while the disk contributes more at short orbital periods and longer wavelengths. Interesting to note is that the disk is more prominent in \C\ than in \D\, despite having a slightly longer orbital period.
    At long periods and with low-SNR data, there can be some degeneracy between the disc and brightspot. For the $r$-band model of \E\ and \B\, no light from the disk is modelled but instead, the brightspot contributed 7\% of the light. We suspect that this is due to minor systemetics in the data and that the 7\% contribution of light is actually from the disk.}
    \label{tab:lumfrac}
\end{table*}

The lightcurves and their best fit models are shown in Figures \ref{fig:lcfit_A}--\ref{fig:lcfit_D},
and the best-fit parameters and uncertainties can be found in the Appendix (Tables \ref{tab:lcpars_A}--\ref{tab:lcpars_D}). An overview of the binary parameters is given in Table \ref{tab:allbinarystats} and Table \ref{tab:lumfrac} lists the contribution of each component to the overall luminosity. 
We first discuss the overall results, and discuss the result for each object individually at the end of this section.

As expected and can be seen in Table \ref{tab:lumfrac}, the relative contribution by the disk and brightspot increases with decreasing orbital periods. In general, for the long period systems (\A, \E, and \B) the brightspot is not detected. However, the $r$-band models for \E\ and \B\ do include a significant contribution by the brightspot. Since this feature is not seen in the other bands, we expect this is due to some asymmetry in the lightcurve caused by residual flickering, and the additional light is actually from the disk. For the two short-period systems (\C\ and \D) there seems to be a detection of the brightspot eclipse. However, this is a subtle feature in the lightcurve. The brightspot eclipse for \D, but only marginally for \C. We note that the contribution from the disk is largest for \C\, despite having a longer orbital period than \D. We discuss this further in Section \ref{subsec:accretionstate}.

Table \ref{tab:allbinarystats} shows the binary parameters as derived from the lightcurve modelling. The main source of uncertainty is the degeneracy is between $q$ and $i$ \citep{green2018a}. For a lightcurve with where only the primary eclipse is detected, $q$ and $i$ are fully degenerate. This degeneracy can be lifted with additional features in the lightcurve; either by measuring the brightspot eclipse, or the eclipse shape which is set by a combination of the limb-darkening of the white dwarf and the white dwarf mass-radius relation used.\footnote{A radial velocity semi-amplitude measurement or surface gravity measurement would also work.} As we have discussed above, only \D\ shows a clear detection of the brightspot eclipse. Therefore, the uncertainty on the mass-ratio is high for the other systems, up to $\approx 50$\% is the worst case (\E). Correspondingly, the uncertainties on the inclinations are 0.5--2 degrees. The uncertainties on the other physical parameters are mostly due to this uncertainty in the inclination. We do note that of the 5 systems, \A\ has be best constrain on the inclination despite only showing a grazing eclipse. This can be explained by the lack of any disk or brightspot and the high quality $g$-band data.

\subsubsection{\A}
The models for \A\ agree well with the data for each of the colour bands and do not show any systematic deviations. This justifies the use of a model with only the accretor and donor, without any contribution from a disk or brightspot. 

The best-fit parameters consistently indicate that the inclination is $82.8\pm1.1$ degrees, which is expected for a grazing eclipse. Inspection of the posteriors shows that $q$ and $i$ are strongly correlated (as expected). The mass-ratio ($q$) is $0.025\pm0.001$. The results are consistent between bands.

The derived binary parameters (masses and radii) are all consistent between bands. The $g$-band results are the most accurate and we use these in the rest of the paper. The mass of the white dwarf is $M_1=0.90\pm0.05$\,\Msun. The donor has a mass of $M_2=0.023$\,\Msun\ and radius of $R_2=0.068\pm0.007$\,\Rsun.

\subsubsection{\E}
The fit to the lightcurves of \E\ are good and no discrepancies can be seen. The best-fit models indicate that almost no light is visible when the system is in eclipse. The parameter estimates for the $g$ and $i$ band are consistent with each-other, but the $r$-band parameters differ. This is possibly because the $r$-band model has a 7\% contribution from the brightspot, while the $g$ and $i$ models attribute any additional light to the disk. This is likely due to some residual systematics in the $r$-band data. Taking this and the data quality into account, we conclude that the $g$-band solution is the most accurate and precise.

As is the case with \A, with just a white dwarf eclipse and no additional information, there is a degeneracy between $i$ and $q$, which are both poorly constrained; $i=86.5\pm2.0$ and $q=0.0214^{+0.016}_{-0.007}$, which results in relatively poor constrains on the mass and radius. The mass of the white dwarf is $M_1=0.79^{+0.13}_{-0.10}$\,\Msun. The donor has a mass of $M_2=0.0165^{+0.018}_{-0.006}$\,\Msun\ and radius of $R_2=0.068\pm0.007$\,\Rsun. 

\subsubsection{\B}
Models for \B\ agree well with the data, and do not show any variability that the models do not account for. We note that in all lightcurves, the eclipse is not consistent with zero, and up to 10\% of the light is emitted by the disk and/or spot. Similar as for \E\, the $r$-band model does model some contribution by the brightspot, but the $g$ and $i$ models does not include any contribution from the brightspot. Again, this is likely due to systematics in the data.  

The parameter estimates between bands deviate significantly. The $g$ and $r$-bands roughly agree, but the $i$-band solution deviates significantly. It is skewed towards very high inclination solutions and very low-mass ratios. A comparison between the models shows a different eclipse shape in the $i$-band compared to the other two models. The difference is small, but does seem to have an effect on the parameter estimates. The difference is possibly a result of the detrending process. Given that the quality of the $g$-data is best, we use this result in the rest of the paper. 

As is the case for \A\ and \E\, the solution for \B\ is also affected by a degeneracy between $i$ and $q$. However, because of the abundance of $g$-band data, the parameters are beter constrained compared to \E\: $i=85.3^{+0.8}_{-0.9}$ and $q=0.018^{+0.007}_{-0.004}$. The mass of the white dwarf is $M_1=0.83^{+0.07}_{-0.08}$\,\Msun. The donor has a mass of $M_2=0.014^{+0.007}_{-0.005}$\,\Msun\ and radius of $R_2=0.054^{+0.008}_{-0.006}$\,\Rsun.

\subsubsection{\C}
This shorter period system shows significant variability out of eclipse. While we did detrend the data and combined a number of orbits, some correlated variability remains in the lightcurve. Besides this, the model is in good agreement with the data. In this system, the disk contribution is clearly visible as the slopes before and after the main eclipse by the white dwarf. There also seems to be some asymmetry present in the eclipse and it seems to be fainter at egress, which is likely the eclipse of the brightspot. This is however at the same level as the residual flickering, and might not be significant.

Estimates of the system parameters show roughly consistent between bands. For sake of consistency, we adopt the $g$-band results in the rest of the paper. Because there is no significant detection of the brightspot eclipse, there is still a strong degeneracy between $q$ and $i$. The inclination and mass-ratio are $i=87.0^{+0.5}_{-1.5}$ and $q=0.030^{+0.002}_{-0.003}$. The mass of the white dwarf is $M_1=0.76^{+0.05}_{-0.04}$\,\Msun. The donor has a mass of $M_2=0.026^{+0.010}_{-0.002}$\,\Msun\ and radius of $R_2=0.049^{+0.006}_{-0.002}$\,\Rsun.

\subsubsection{\D}
This system has an orbital period very similar to \C, but does not show as much flickering and has a smaller contribution to the luminosity by the disk (Table \ref{tab:lumfrac}). The contribution by the brightspot is more significant and can be seen by the broad 'hump' which peaks at phase $\approx0.8$. In addition, there is also a possible contribution by a `superhump' that causes a peak in the lightcurve at phase 0.2 (also seen in YZ Lmi, see \citealt{copperwheat2011}). In this system, the brightspot eclipse is clearly visible (at phase 0.02--0.08) as shown by the inset.

 The parameter estimates from the $g$ and $i$ band both agree, but the $r$-band result indicates a lower inclination and higher mass-ratio. There is no obvious difference between lightcurves. Comparing the models, there is a subtle difference between the duration of the eclipse between bands. This could be the result of the flickering and/or the detrending process. We again adopt the $g$-band solution for the rest of the paper, as it is deemed the most reliable given the highest SNR and largest amount of data available.

The inclination and mass-ratio are $i=86.5^{+0.8}_{-0.5}$ and $q=0.024^{+0.003}_{-0.004}$. The mass of the white dwarf is $M_1=0.79^{+0.04}_{-0.05}$\,\Msun. The donor has a mass of $M_2=0.019^{+0.003}_{-0.004}$\,\Msun\ and radius of $R_2=0.044^{+0.002}_{-0.003}$\,\Rsun.

\subsection{The average spectra and spectral lines}

\begin{figure*}
\includegraphics[width=\textwidth]{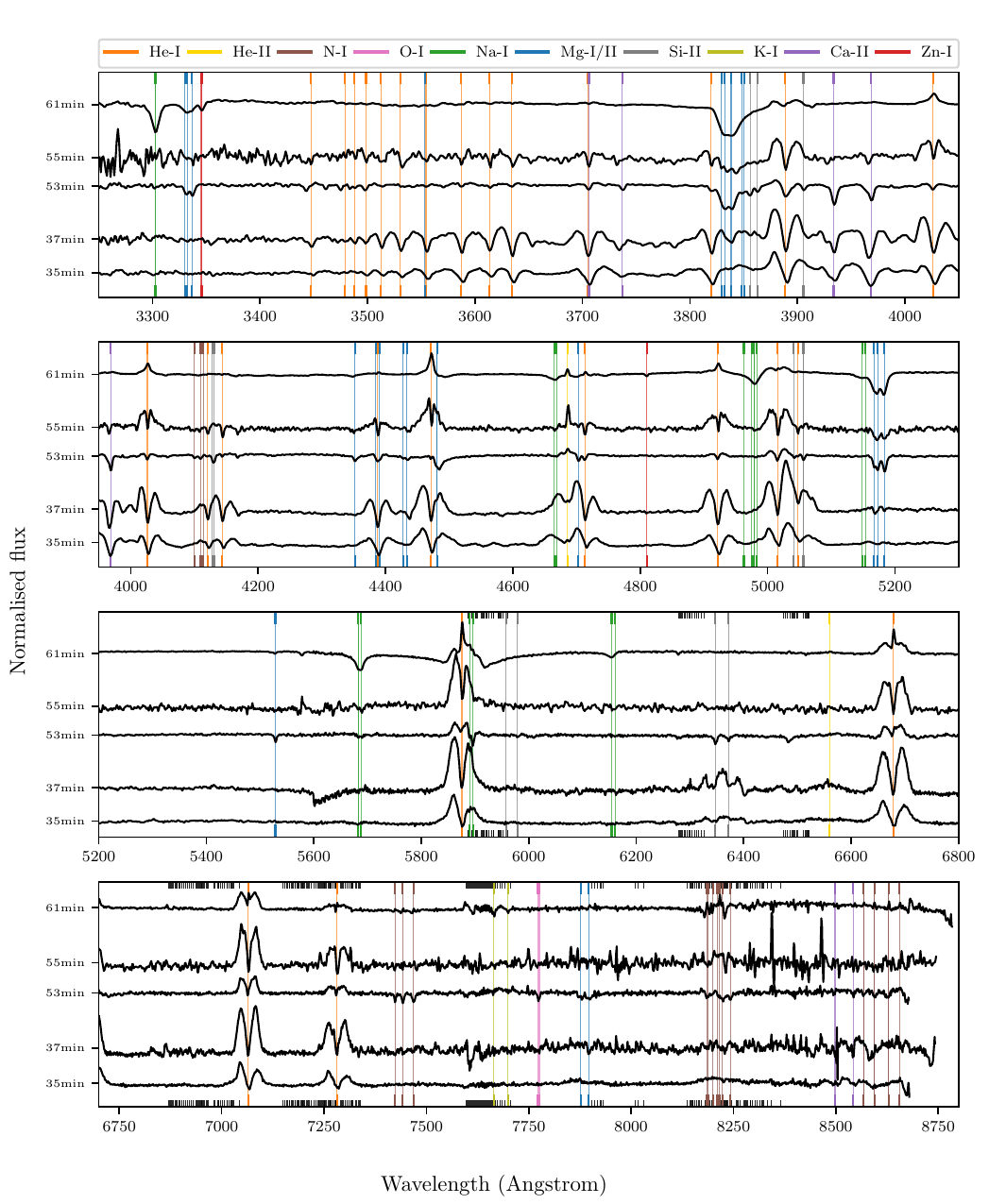}
\caption{The averaged and normalised LRIS spectra of \A, \E, \B, \C, and \D\ from top to bottom, labelled by their orbital periods. The spectra have only been shifted and have not been re-scaled. The different absorption and emission lines are indicated by coloured vertical lines, and telluric lines are indicated with black ticks at the top and bottom.}
\label{fig:avgspec}
\end{figure*}

\begin{table}
    \centering
    \begin{tabular}{l|ccccc}
& \rot{\A}\ & \rot{\E}\ & \rot{\B}\ & \rot{\C}\ & \rot{\D}\ \\
\hline
\hline
    He-\textsc{I}  & \gb{}  e, de & \gb{}e  & \gb{}de  & \gb{}de & \gb{}de \\
    He-\textsc{II}  &\gb{}e & \gb{}e &\gb{} e &\gb{} de &\gb{} de\\
    N-\textsc{I}  & \gb{}e, a & & \gb{}a & \gb{}a? & \gb{}a? \\
    O-\textsc{I}  &  & & \gb{}a  & &  \\
    Na-\textsc{I}  & \gb{}a, \gb{}e? & & \gb{}a & \gb{}a & \gb{}a?\\
    Mg-\textsc{I}  & \gb{}a & \gb{}a & \gb{}a & \gb{}a & \gb{}a? \\
    Mg-\textsc{II}  & & & \gb{}a &  & \gb{}a? \\
    Si-\textsc{I}  & & & \gb{}a & \gb{}de & \gb{}de \\
    K-\textsc{I}  & \gb{}a &&  & & \\
    Ca-\textsc{II}  & & & \gb{}a & \gb{}a & \gb{}a \\
    Zn-\textsc{I}  & \gb{}a & & & & \\
    \hline
    \end{tabular}
    \caption{List of elements identified from emission and absorption lines in the averaged LRIS spectra (see Figure \ref{fig:avgspec}). `e' indicates emission lines, `de' indicates double-peaked emission lines, and `a' for absorption lines.}
    \label{tab:lines}
\end{table}

Spectra of AM~CVn systems are a combination of light from the white dwarf, the disk, the brightspot, and sometimes a boundary layer on the interface between the disk and white dwarf \citep[e.g.][]{kupfer2015}. The donor is cold and not visible. For systems in quiescence, the overall shape of the spectral energy distribution is set by the continuum of the white dwarf. The brightspot is small but hot and is sometimes seen as excess UV emission.


In quiescence, the disk is optically thin and appears as emission lines. If the system is viewed near edge-on, the lines are double-peaked. 
In GP Com \citep{smak1975,morales-rueda2003,kupfer2016} and other systems, a single-peaked emission line is also visible in the centre of many double-peaked He lines, dubbed the `central spike'.  Radial velocity measurements show that these lines originate from the white dwarf \citep{marsh1999,roelofs2005,kupfer2015} and is likely boundary layer emission from the surface of the white dwarf \citep[e.g.][]{green2019,gehron2014}.

In some cases, absorption lines of metals are present in the spectra of AM~CVn systems \citep{groot2001,anderson2008,kupfer2016}. Radial velocity shifts of absorption lines are consistent with that of the white dwarf \citep{kupfer2016,green2020}, which confirms that these lines are formed in the atmosphere of the white dwarf.

The average spectra are shown in Figure  \ref{fig:avgspec}, and the spectral lines are listed in Table \ref{tab:lines}. To identify the lines, we systematically searched the spectra element by element and checked for the presence of spectral lines. First, we confirmed that no Hydrogen lines are present. We continued with He-\textsc{I} lines, which are found both in emission (as double peaks) and absorption in all spectra. \A\ also shows a strong `central spike' component. 
He-\textsc{II} is also present in all spectra as a small but significant emission line, which is also likely a `central spike'.  There also seems to be some double-peaked emission from He-\textsc{II}, but other lines close to He-\textsc{II}-4686\AA\ makes this hard to determine with certainty. \E\ is an exception and shows a strong emission line at He-\textsc{II}-4686\AA.

We continued the search with the CNO elements. N-\textsc{I} lines are in emission in the spectra of \A\ and absorption for \B. The spectrum of \D\ seems to show a broad emission feature consistent with emission from the N-\textsc{I} multiplet at 8223\AA\ \citep[for an overview of N lines see][]{kupfer2017}. There are also some hints of emission lines by N-\textsc{I} around 8680\AA. There is no clear detection of N-\textsc{I} in the spectra of \E\ and \C , possibly because the SNR of these spectra a lower.
Oxygen is clearly detected in \B: there is a clear detection of an absorption feature at 7772/4/5 \AA. This feature is not detected in the other spectra. We found no carbon lines in any of the spectra, which has implications on the nature of the donor star, see Section \ref{sec:donorproperties}.

Next, we searched for the presence of metal lines in the spectra. Both Na-\textsc{I} and Mg-\textsc{I/II} are detected as absorption lines. Mg-\textsc{I} can be seen in the triplets at 3832\AA\ and 5172\AA. They are strongest for the long period system and decrease in strength for the short-period systems. Mg-\textsc{II} lines are weaker but can be seen at 4481\AA\ and possibly at 4384/90. Na-\textsc{I} lines are especially strong in \A, but barely detectable in the other spectra (3302/3, 5682/8, 6154/60, and 5889/95\AA). Ca-\textsc{II} lines (3933/68\AA\ and also 3706/36\AA) are seen for \E, \B, \C, and \D, but not for \A.
Finally, \B, \C, and \D\ all show Si-\textsc{II} lines in their spectra; as double-peaked emission lines (6347/71\AA) for the two short-period systems, and as absorption lines for \B.  There are also lines visible at 3856/62\AA\ for \B\ and \C. No Si-\textsc{II} is visible for \A, and \E. 

There are a few rare elements present. In \A, two Zn-\textsc{I} absorption lines can be seen (3345\AA\ and 4810\AA). In addition, there are two lines present (7665/95\AA) in \A\ which are consistent with the strongest two K-\textsc{I} lines. We do note that these are located close to telluric absorption lines.

\section{Discussion}\label{sec:discussion}

\subsection{Search for eclipsing AM~CVn binaries}
\subsubsection{Recovery efficiency and space density}
\gaia\ has measured a parallax for all 7 eclipsing AM~CVn systems, and the distance is known with a precision of 5\% to 30\%.
With a sample size of 5 (and 2 previously known eclipsing AM~CVn systems), only an order of magnitude estimate of the local space density is possible. However, there is a two order of magnitude difference between the predicted AM~CVn space density \citep{nelemans2001,kremer2017} and the measured space density by \citet{carter2013}, and it is therefore useful to check if the yield is consistent with the measured space density to exclude the possibility of a previously missed population. In addition, an estimate of the recovery efficiency is useful to estimate how many systems ZTF and other surveys are set to discover in the near future.

We have done a very thorough search of objects in the white dwarfs catalogue by \citet{gentilefusillo2019}, and limit the discussion to this sample. Of the 7 known eclipsing AM~CVn systems, only \C\ is not in this catalogue, and we therefore exclude it in this estimate. Both already known eclipsing AM~CVn systems, Gaia14aae and YZ Lmi, are part of this catalogue. We recovered Gaia14aae in our search, but we initially did not recover YZ LMi because its orbital period is just short of the initial cutoff of 28.8 minutes (0.02 days). Therefore, we searched all lightcurves with one or more alerts again, but now down to periods of 14.4 minutes (0.01 days) days and did recover YZ LMi with the correct period, but no other systems were found.

A total of 241\,775 objects in this sample have ZTF lightcurves with more than 80 epochs (49.6\% of the total catalogue). To estimate the recovery efficiency for each object, we assume that if there are 5--7 or more in-eclipse points in the lightcurve, we are able to identify the source as an eclipsing AM~CVn system. This can be easily calculated as it is just a function of the total number of epochs in the lightcurve and the eclipse duty cycle (see Appendix \ref{app:recovery}). 

We estimate the eclipse duty-cycle for each system and calculate the average probability to recover them from the 241\,775 ZTF lightcurves. The average recovery efficiencies are 33\% for \A, and  60--75\% for the other systems (including Gaia14aae and YZ Lmi). This suggests that there are $\approx$ 1--4 eclipsing systems we have not detected which do have ZTF lightcurves and another 6--10 outside of the ZTF footprint.   

In order to estimate the space density, we use the basic $\mathrm{1/V_{max}}$ method \citep{schmidt1968}, with distances from \citet{bailer-jones2021} (see Table \ref{tab:overview}). Since the AM~CVn systems are very close, we do not correct for expected changes in the space density and assume a uniform distribution.
We use the \gaia\ parallax and a magnitude limit of 20.5 to estimate the total volume each of the systems could have been recovered. We correct this estimate by the recovery efficiency we estimated earlier, a correction for the fraction of systems not in the ZTF footprint, and a correction factor of 8 (the average correction factor to account for eclipsing Roche-lobe filling binaries with mass-ratios between $0.01<q<0.05$). This results in an estimated space density of $\rho=6^{+6}_{-2} \times 10^{-7}$\,$\mathrm{pc^{-3}}$. This consistent with the measured space density by \citet{carter2013}, $\rho=5\pm3 \times 10^{-7}$\,$\mathrm{pc^{-3}}$ and again confirms the discrepancy with the population synthesis predictions \citep[e.g.][]{nelemans2001}. With this search, we exclude the possibility that this discrepancy is due to a large hidden population of faint, low-accretion rate AM~CVn systems.

\subsubsection{Observational properties compared with the known population of AM~CVn systems}\label{sec:obsprops}

The currently known sample of AM~CVn binaries has been found by various searches based on static colours, outbursts, short periods variability, or by large spectroscopic surveys. These search methods have different inherent biases. Here, we briefly discuss the biases of each method, and compare them to the search for AM~CVn system using eclipses. 

Searches for outbursting AM~CVn \citep[e.g.][]{levitan2015} will predominantly find systems with periods between 22 and $\unsim50$ minutes. A search based on SDSS colours by \citet{carter2013} has mostly found AM~CVn systems with intermediate periods as well. These systems stand out because their colour is a combination of a DB white dwarf and some contribution by a disk. Shorter period systems (direct impact accretion systems) are easy to find because of their high X-ray luminosity but these are intrinsically rare due to the fast evolution. Large scale spectroscopic surveys  allow for the discovery of systems at all periods \citep[e.g.][]{anderson2005}. Therefore, most of the long period ($\gtrsim 50$\,min) systems have been discovered using SDSS spectra.

A search for AM~CVn systems by their eclipses has very different biases. 
It is biased towards high inclination systems. Second, systems with deep eclipses are easier to detect than shallow eclipses (assuming an efficient pipeline that can handle `dropout'). Eclipses are more shallow for short period systems where the disk or brightspot dominate the luminosity. However, short period systems can show strong superhump periods which are close to the orbital period. If outbursts and flickering dominate the lightcurve, the period is more difficult to determine, which is an additional complicating factor.
Eclipsing long period systems are easy to find; the white dwarf dominates the luminosity which means the eclipses are deep. In conclusion, using eclipses to find AM~CVn systems is biased to low accretion rate and therefore long orbital periods. The eclipse probability does not decrease significantly with the orbital period since the donor size increases with the orbital period. 

\begin{figure*}
  \centering
  \subfigure[The \gaia\ observational H-R diagram (eDR3) with as background stars within 200\,pc.]{\includegraphics[scale=0.95]{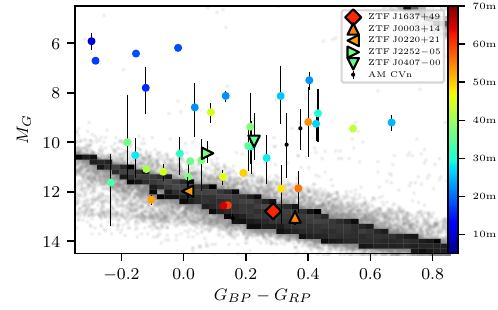}} \hspace{3mm}
  \subfigure[The SDSS $u-g$ versus $g-r$, colour-colour diagram of the white dwarf catalog \citep{gentilefusillo2019}. The red line shows the colour-selection used by \citet{carter2013}. ] {\includegraphics[scale=0.95]{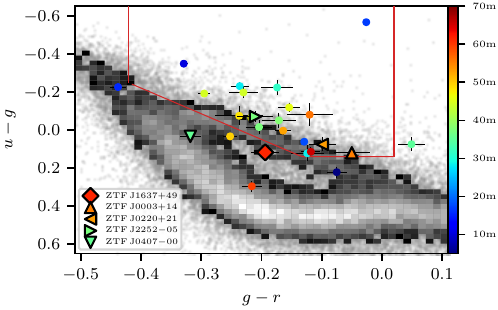} }
  \caption{The new eclipsing systems compared to known AM~CVn systems in the HR-diagram and colour-colour diagram. The newly discovered eclipsing systems are shown as triangles and a diamond and the known AM~CVn systems as dots. The marker colour indicates the orbital period of the system. Note that we corrected the SDSS measurements for \A\ and \E\ that were obtained in-eclipse. In Section \ref{sec:obsprops} we discuss why these systems were not identified as AM~CVn systems earlier and the implications for future searches based on colours and parallaxes.}
  \label{fig:population}
\end{figure*}

To determine why none of these eclipsing systems have been identified earlier as AM~CVn systems, we plot the detected eclipsing AM~CVn systems and the known sample in an HR diagram and a $u$-$g$--$g$-$r$ colour diagram (Figure \ref{fig:population}). 

\A\ and \B\ are located on the white dwarf track in the HR diagram, and close to the DB track in colour-space. Both of these systems are dominated by the DB white dwarf, and light emitted from the disk or brightspot is almost negligible. Both these systems appear as typical DB white dwarfs.
\E\ is very red and an outlier in both figures. This could be due to reddening ($E_{g-r}=0.14$). In the HR-diagram, it is located slightly above the white dwarf locus, which is expected given the small contribution from a disk and the low temperature. \C\ and \D\ are located above the white dwarf track in the HR diagram which suggests that the contribution by the disk and the brightspot was small but not negligible when they were observed by \gaia. 

In conclusion, finding short period, non-eclipsing systems like \C\ and \D\ is easier because they stand out from the white dwarfs in the HR diagram. In addition, they also show outbursts and flickering. However, systems like \A, \E, \B\ do not stand out in either diagram and also do not show any outburst. If they are not eclipsing, the only solution is to obtain spectra of all white dwarfs and identify them by the He emission lines. Upcoming large spectroscopic surveys like SDSS-V \citep{kollmeier2017}, 4MOST \citep{dejong2012}, and WEAVE \citep{dalton2012,dalton2014,dalton2016}, have programs to observe white dwarfs and will be able to find long period AM CVn systems that are non-eclipsing.

We note that all 5 systems are in the SDSS footprint and have $ugriz$-colours, although none of them was found in the search for AM~CVn systems using SDSS-colours by \citet{carter2013} and \citet{carter2014}. \A\ was observed during the eclipse which mostly affected the $r$-band measurement and caused the system to be rejected as a candidate. \E\ was in outburst when SDSS observed it, and the colours fall outside of the selection criteria (we used the Pan-STARRS $g$ and $r$ measurements in Figure \ref{fig:population}). The primary SDSS measurement of \D\ also falls outside of the selection. However, it has been observed 14 times by SDSS, and only 2 out of these 14 do not pass the selection criteria. \B\ and \C\ both pass the selection criteria, but no spectra were obtained; the spectroscopic completeness of \citet{carter2013} is 70\%. While the fact that none of these systems were found by \citet{carter2013} initially suggested that they underestimated their recovery efficiency, but there is a plausible explanation for why all of these systems were not discovered. We conclude that the space density estimate by \citet{carter2013} is not significantly affected by any systematic bias or uncertainty.

\subsection{Outbursts and accretion state}\label{subsec:accretionstate}
The overall trend for AM~CVn systems is that the average accretion rate decreases with the orbital period. At orbital periods of 22 minutes and longer, the disk is generally in a low state with outbursts semi-regularly. \citet{levitan2015} showed that there is an exponential correlation ($\tau \propto P_\mathrm{orb}^{7.35}$) between the orbital period and outburst frequency, with a super-outburst frequency of one year at an orbital period of 34 minutes. The behaviour of the systems we found show behaviour consistent with this correlation, with a few detected outbursts for the two short-period systems over the last decade, and no detected outbursts for the longer period systems.

Because these systems are eclipsing we can determine the relative luminosity of the disk and because the inclinations are similar also directly compare the spectra. Lightcurve modelling shows that the relative contribution by the disk is $\approx 10$--$30\%$ for the short period systems, just $\approx 5\%$ for the 50 minute systems, and seems to be non-existent for the longest period system (see Table \ref{tab:lumfrac}). A useful aspect of our sample is that there are two pairs of systems with almost identical periods. A direct comparison between the systems in each pair shows interesting differences. As can be seen in both the lightcurves and the spectra  (Figure  \ref{fig:avgspec}), the contribution by the disk varies significantly between the members of each pair. This is possibly due to the recent outburst, which would empty the disk and heat the white dwarf, but could also be due to the more compact nature of the donor. Continued monitoring of the systems using high-speed photometry and high SNR lightcurves to more precisely measure the mass and radius of the donor \citep{green2018} will enable us to resolve these issues.

For \D, we observe one outburst in the ZTF data, which showed multiple short 'echo' outbursts days to weeks after the main superoutburst, also seen by \citet{green2020}, \citet{duffy2021}, and \citet{pichardomarcano2021}. Well sampled lightcurves of the systems also show a brighter baseline level and up to 20 echo-outbursts after a superoutburst. In the case of \D, we also observe a change in the eclipse depth; the eclipse is total in the ZTF data (see Figure \ref{fig:ZTFlcs}) but only 60\% deep in the Chimera data. These three observations suggest a change in the state of the system, possibly an increased temperature of the white dwarf, a brighter disk, or both.

Because \D\ is eclipsing, we can potentially measure the changes in the systems more directly using the eclipses. 
By obtaining high-speed photometry before, during, and after a superoutburst, we can measure changes both in the white dwarf temperature as well as the size and brightness of the disk \citep[e.g.][]{copperwheat2011}. 

\begin{figure*}
	\includegraphics[width=\textwidth]{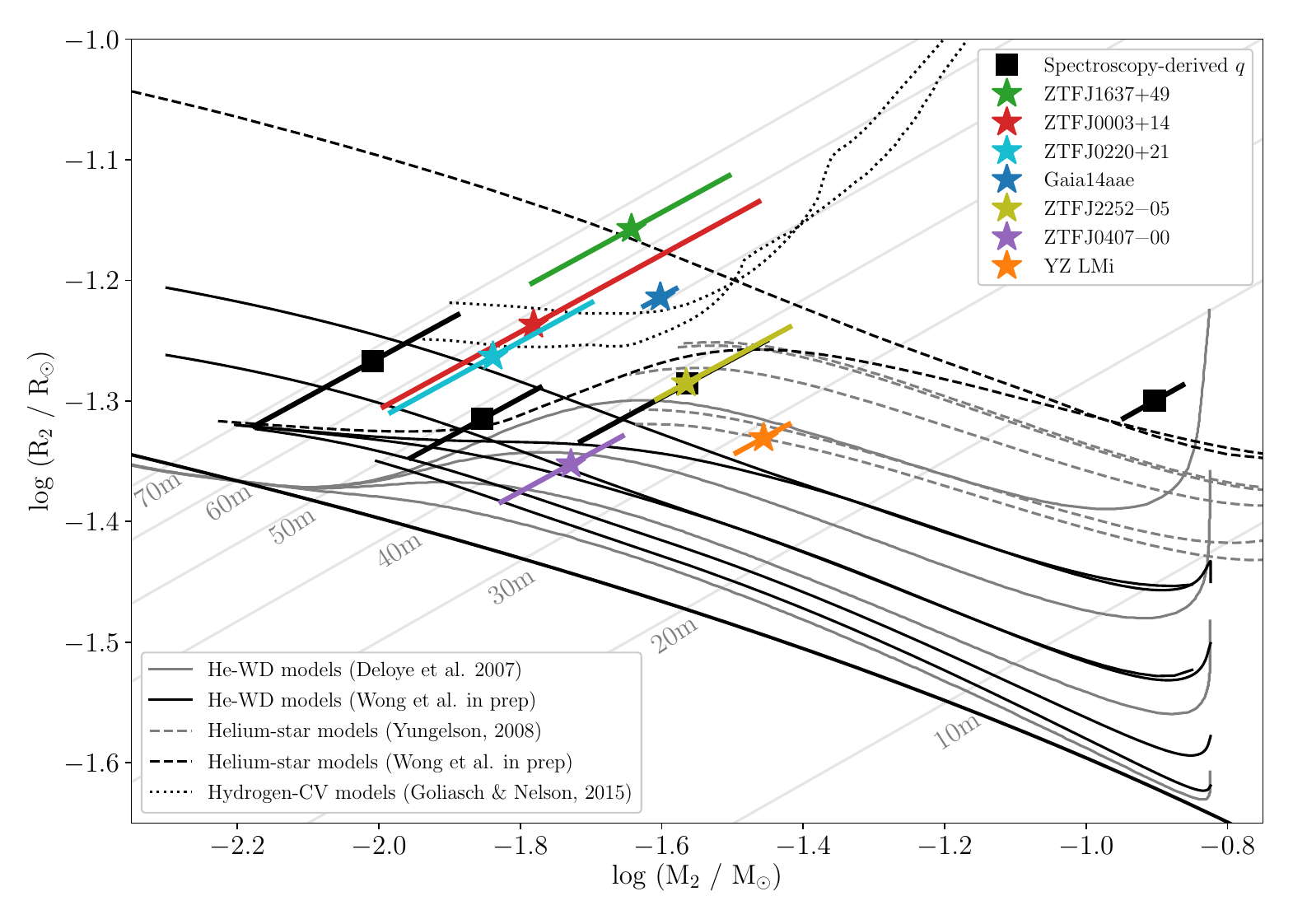}
    \caption{The mass and radius of donor stars in AM~CVn systems compared to model predictions, adapted from \citet{green2018a}. Stars indicate measurements of eclipsing systems, the 5 new systems from this paper, YZ LMi from \citet{copperwheat2011}, and Gaia14aae from \citet{green2018}. Solid lines show the white dwarf channel models (\citealt{deloye2007} in grey, Wong et al. in prep in black), dashed lines show the He-star channel models (\citealt{yungelson2008} in grey, Wong et al. in prep. in black), and dotted lines show the evolved CV channel models \citep{goliasch2015}. The white dwarf and helium star tracks include multiple levels of entropy, with the lower entropy models at the bottom. The lowest line shows a zero-entropy donor. Models by Wong et al. diverge; donors that are allowed to cool shrink at long orbital periods, while donors that are not allowed to cool and lose entropy as they lose mass stay large.}
    \label{fig:MR}
\end{figure*}

Besides the usual disk emission line, we note that a small but significant He-II--4686\AA\ central spike is detected. A high temperature is required to ionise helium, which suggests that there is a hot, but small boundary layer in all five systems. This feature is also seen in other long-period systems \citep[e.g.][]{kupfer2015} and speculate that this feature is present to some extent in all AM~CVn systems, but they can only be detected with high SNR spectra.

\subsection{The properties of the donor stars}\label{sec:donorproperties}
\subsubsection{Chemical composition}
The spectra show many different spectral lines, see Figure  \ref{fig:avgspec} and Table \ref{tab:lines}. Modelling the spectra to determine elemental abundances is challenging and beyond the scope of this paper. We limit this discussion to qualitative estimates only.


\citet{nelemans2010} showed that abundances can be used to determine the evolutionary channel the donor formed through. First, the presence of any hydrogen unambiguously points to an evolved main-sequence donor \citep[e.g.][]{breedt2014,green2020}. We do not detect any hydrogen in the spectra of any of the systems. Since even a trace amount of hydrogen should be visible in the spectra, we conclude that none of the systems are formed through the 'evolved' CV channel.

When the donors have experienced significant helium burning, as is the case for helium star donors, the N abundance decreases to almost undetectable levels. Detailed helium star models for the donor stars in AM~CVn systems predict typical surface abundance ratios N/C=0.001--1 while for white-dwarf donor stars N/C$\approx0.1$ if no CNO burning took place, or N/C$>>1$ if CNO burning did take place.

We have detected N in both \A\ and \B\ and do not detect any C. This is typical for many AM~CVn systems and suggests that these systems are formed through the He-WD model. 

In \B\ we also detect oxygen, which suggests a O/N$\approx1$. The O/N ratio in the WD model is set by the progenitor mass. Models by \citet{nelemans2010} predict a ratio of O/N of close to 1 for a 1\,\Msun\ progenitor mass, and a O/N ratio of 0.1 for a 2\,\Msun\ initial mass. This suggests that \A\ evolved from WD donor with a higher initial mass, while \B\ evolved from a lower initial mass WD donor.

For systems \E, \C, and \D, there is no clear detection of N lines, which makes any inference of the formation channel impossible. 

Besides the CNO lines, we also observe a variety of metal absorption lines. The line strengths depend on the element in question and a combination of sedimentation timescales \citet{koester2009}, temperature, surface gravity, accretion rate, and abundances of the accreted material. 

The expectation is that the abundances of metals are primordial. However, sedimentation and mixing during the donor star lifetime affect the current abundance of metals. We have measured the temperature and surface gravity of the white dwarf, and by modelling the spectra using atmosphere models, we can in principle constrain the abundance of the metals but this is beyond the scope of this paper. This can shed light on both the primordial abundance in these systems as well as constrain the amount of sedimentation and/or mixing that took place during the formation of these systems.



\begin{figure}
  \centering
  \subfigure[The absolute $g$-band magnitude versus the $g$-$r$ colour of the white dwarf in eclipsing AM~CVn systems. The measurements have been de-reddened based on the $g-r$ reddening (see Table \ref{tab:overview}), indicated with lines with the same colour as the markers. Model temperature are indicated below the model in kK.]{\includegraphics[width=0.99\columnwidth]{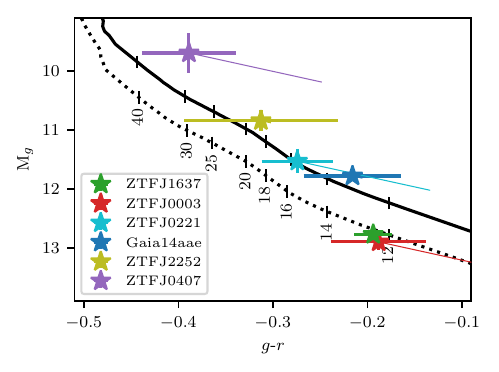}} \hspace{3mm}
  \subfigure[The temperature and mass of the white dwarf in eclipsing AM~CVn systems versus the orbital period. The horizontal lines in the bottom panels show the average mass and the one standard deviation range. ] {\includegraphics[width=0.99\columnwidth]{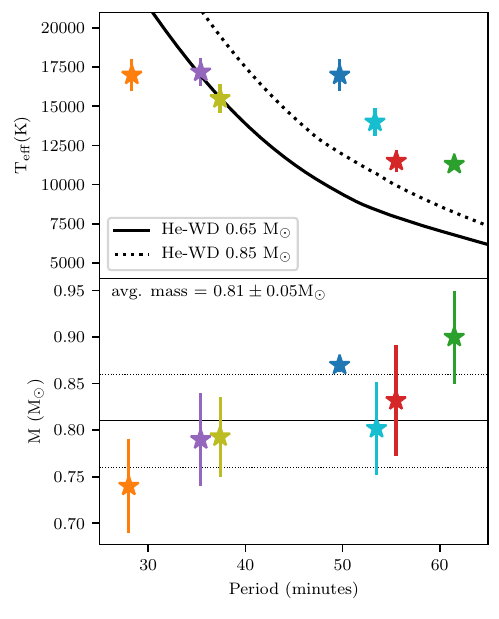} }
  \caption{Properties of the accreting white dwarfs in eclipsing AM~CVn systems. Models for a 0.65\,\Msun\ (dashed line) and 0.85\,\Msun\ (solid line) white dwarf and are from Wong et al. in prep. Values for YZ LMi and Gaia14aae have been taken from \citet{copperwheat2011} and \citet{green2018} and \citet{green2019a}.}
  \label{fig:WDprops}
\end{figure}

\subsubsection{Mass and radius}
Figure \ref{fig:MR} shows the mass and radius of the donors in eclipsing AM~CVn systems and four AM~CVn systems for which the mass-ratio has been measured using spectroscopy \citep{green2018}. Because the donor is filling its Roche-lobe, the density of the donor depends only on the orbital period, indicated by diagonal lines in the figure \citep{faulkner1972,green2020}. Donors lose mass as the orbital period of AM~CVn systems increase, and evolve from right to left in this diagram. The WD and He-star donors are degenerate, and they increase in size as they lose mass.

The donor-models for each formation channel predict a different entropy, and therefore a different mass and radius for a given period. In the figure, we show tracks for evolved CVs \citep{goliasch2015}, the white dwarf track \citet{deloye2007}, and tracks for the He-star channel from \citep{yungelson2008}. We also show WD-tracks and He-star tracks by Wong et al. in prep. The tracks by Wong et al. diverge; the models that curve downwards include cooling, the tracks that continue on to the top-left do not include cooling. 

First of all, the measurements indicate that masses are 1.5--2.5 times more massive than a zero-entropy donor for the same orbital period. This is similar to what has been measured for YZ LMi and Gaia14aae. 

The short-period systems seem to be consistent with either white dwarf models or He-star models. However, current WD and He-star models predict that the donors cool and shrink at orbital periods of 30-40 minutes, which is not consistent with the observations. 

The long period systems seem to be very large, similar to Gaia14aae. As noted by \citet{green2018}, the measurements are consistent with CV-channel donors, but this channel is excluded by the lack of hydrogen in the spectra \citep{nelemans2010}. In addition, this channel is also predicted to be the least common, although more recent work by \citet{liu2021} suggest this might not be the case. 

The alternative solution is that AM~CVn systems do evolve through the WD and/or He-star channel but for some reason do not cool and shrink at long orbital periods (indicated by the models by Wong et al.). To answer this question, the donor properties of the new ZTF systems need to be measured with similar accuracy as for Gaia14aae. In addition, precise characterization of donors of to-be discovered short period eclipsing systems ($\lesssim 25$\,min) are needed to constrain the models before adiabatic cooling is predicted to become important. In addition, models need to be updated to make sure any source of potential heating of the donor is taken into account.

\subsection{Properties of the white dwarf}\label{subsec:WDprops}


Figure \ref{fig:WDprops} shows the properties of the white dwarf in eclipsing AM~CVn systems. We first compare the absolute magnitude versus the colour of the white dwarf with model values. We calculate these values by using the Pan-STARRS measurement, \gaia\ eDR3 distances, the estimated white dwarf contribution from table \ref{tab:lumfrac}, and the reddening towards the system (see table \ref{tab:overview}). A general prediction is that the effective temperature of white dwarfs in AM~CVn systems are determined by the average accretion rate and therefore correlated with the orbital period \citep{bildsten2006}. Our measurements generally agree with the model values calculated by Wong et al. (in prep.).
Although the errorbars are too large to estimate the mass, the general trend is that the long period systems are massive $\unsim 0.85$\,\Msun, while the shorter period systems seem more consistent with lower masses $\unsim 0.65$\,\Msun. 

We have measured the temperature from the SEDs and the mass using the eclipses. As we have already seen from the colour of the white dwarf, long-period systems are colder while short period systems contain a hotter white dwarf However, the temperature for the longer period systems seem systematically larger than models predict at those particular periods.

This could be due to a larger than expected donor entropy which results in a higher mass-accretion rate and therefore higher temperature. However, it could also be due to an additional source of energy in the white dwarf that causes the white dwarf to remain hot for a longer time \citep[e.g.][]{bauer2020}.
It is possible that the temperature estimates of the short period systems are systematically underestimated, possibly by imperfect corrections to the overall luminosity. Alternatively, metals in the white dwarfs atmosphere can significantly suppress the UV-luminosity, which results in an underestimate of the temperature. Since the SED of hot white dwarfs peaks in the UV, short-period systems would be more affected.

By modelling the eclipse lightcurves and assuming a white dwarf M-R relation we have also determined the mass of the white dwarfs. These are similar to the white dwarf masses in the other eclipsing AM~CVn systems (Gaia14aae and YZ Lmi). The unweighted average and variance of the mass is $0.81\pm0.05$\,\Msun. 
This is more massive than the typical isolated white dwarfs \citep{kepler2007}, and similar to white dwarfs in hydrogen-rich CVs which have an average mass of $\approx0.83$\,\Msun\ \citep{littlefair2008,savoury2011,pala2017}.

However, the explanation for this is different. For hydrogen CVs, angular momentum loss causes systems with low-mass white dwarfs to merge during the early stages of the CV phase \citep{schreiber2016,nelemans2016, zorotovic2020}. In the case of AM~CVn systems, accretion in systems with low-mass white dwarf is unstable and causes a merger during the common-envelope \citep[e.g.][]{shen2015}. 

\citet{kilic2016} show that the number of known white dwarfs with $M\gtrsim0.8$\,\Msun\ and low mass white dwarf companions (WD-channel) is sufficient to explain the AM~CVn birth rate. However, the low-mass white dwarfs have a low entropy which would result in cool and small donors \citep{deloye2007}, inconsistent with our results (Figure \ref{fig:MR}).

\section{Conclusion and summary}\label{sec:summary}


We searched the ZTF lightcurves for deep eclipsing white dwarfs and identified five new eclipsing AM~CVn systems with periods ranging from 35 to 62 minutes. Using this sample, we estimated the local space density which is consistent with previous observational estimates. This is again two orders of magnitude less than model predictions and excludes the possibility that this is due to a hidden population of faint, long period AM~CVn systems. 

We obtained phase-resolved spectra and high cadence lightcurves to characterize the systems. The high SNR averaged spectra of the longer period systems revealed many broad metal absorption lines, including potassium and zinc which have not previously been detected in any other AM~CVn systems. Doppler maps show the presence of a 'central spike' for the longest period system. 

We modelled the high-candence lightcurves and the spectral energy distributions and measured the binary parameters (masses, radii, inclination, and white dwarf temperature) for all five systems. First, we showed that the effective accreting white dwarf temperatures of long-period systems are higher than models predict. This suggests a delay in cooling of the white dwarf. Second, the average accreting white dwarf mass is $\approx 0.8$\,\Msun, more massive than typical single white dwarfs. This suggests that AM~CVn systems can only be formed with massive accreting white dwarfs. Third, the donor stars have a high entropy, and are a factor of 1.5-2.5 times more massive than a zero-temperature donor at the same orbital period. The high observed entropy (radius) is consistent with white dwarf or He-star models for the two short-period systems. The long period systems also have a large entropy (radius), while both WD and He models predict that AM~CVn donor should decrease in entropy (radius) as they evolve to periods longer than $\approx 40$\,min. More accurate donor mass-radius measurements and more complete models are needed to resolve this inconsistency.

\section{Future work}\label{sec:futurework}
We will continue to obtain data on the systems presented in this paper to fully utilize their potential. We aim to obtain high SNR, high-cadence lightcurves in order to measure the donor mass and radius to a precision of $\unsim$1\% (similar to that of Gaia14aae). Continued monitoring of these systems with high-speed cameras will also allow us to measure deviations in the eclipse arrival times.



As estimated in this paper, we expect there to be another handful of eclipsing AM~CVn systems brighter than $\approx\!$20 mag. We will continue to search for new ZTF data for more eclipsing systems. Other survey telescope lightcurves (ATLAS, \gaia, BlackGEM, and \citealt{groot2019}) can also be used to find eclipsing AM~CVn systems. This does require the ability to detect deep eclipses, either by implementing image differencing or by using forced photometry. 

The Vera C. Rubin observatory \citep{ive2019} is starting observations in the near future. With a single epoch limiting magnitude of \mbox{$r\approx24$}, it will observe thousands of AM~CVn systems. However, spectroscopic verification spectra of this faint population is expensive, and identifying AM~CVn systems by measuring the period using eclipses can be done with just the \LSST\ data. This will limit the identification to eclipsing AM~CVn systems only, but there should be tens to a hundred of these in the \LSST\ footprint. Such a large sample of eclipsing AM~CVn systems will provide a much clearer picture of the donor properties and with that the formation channels of AM~CVn systems. We have shown that at least 5 in-eclipse measurements are needed to identify a period, which implies that a total of hundreds of epochs will be needed (see \ref{app:recovery}). The \LSST\ cadence will be sparse compared to ZTF and it will take $>5$ years before enough epochs will be collected to enable a systematic search.

\section*{Acknowledgements}
JvR is partially supported by NASA-\LISA\ grant 80NSSC19K0325.
MJG was supported by the European Research Council (ERC) under the European Union's FP7 Programme, Grant No. 833031 (PI: Dan Maoz). V.S.D and HiPERCAM are supported by STFC. 

This research benefited from interactions at the ZTF Theory Network Meeting that were funded by the Gordon and Betty Moore Foundation through Grant GBMF5076 and support from the National Science Foundation through PHY-1748958. 

Based on observations obtained with the Samuel Oschin Telescope 48-inch and the 60-inch Telescope at the Palomar Observatory as part of the Zwicky Transient Facility project. ZTF is supported by the National Science Foundation under Grant No. AST-1440341 and a collaboration including Caltech, IPAC, the Weizmann Institute for Science, the Oskar Klein Center at Stockholm University, the University of Maryland, the University of Washington, Deutsches Elektronen-Synchrotron and Humboldt University, Los Alamos National Laboratories, the TANGO Consortium of Taiwan, the University of Wisconsin at Milwaukee, and Lawrence Berkeley National Laboratories. Operations are conducted by COO, IPAC, and UW.

Some of the data presented herein were obtained at the W.M. Keck Observatory, which is operated as a scientific partnership among the California Institute of Technology, the University of California and the National Aeronautics and Space Administration. The Observatory was made possible by the generous financial support of the W.M. Keck Foundation. The authors wish to recognize and acknowledge the very significant cultural role and reverence that the summit of Mauna Kea has always had within the indigenous Hawaiian community. We are most fortunate to have the opportunity to conduct observations from this mountain.

The research leading to these results has received funding from the European Research Council under the European Union's Horizon 2020 research and innovation programme numbers 677706 (WD3D) and 340040 (HiPERCAM).

Based on observations made with the Gran Telescopio Canarias (GTC) installed in the Spanish Observatorio del Roque de los Muchachos of the Instituto de Astrof\'isica de Canarias, in the island of La Palma.

Based on observations obtained at the international Gemini Observatory, a program of NSF's NOIRLab, which is managed by the Association of Universities for Research in Astronomy (AURA) under a cooperative agreement with the National Science Foundation on behalf of the Gemini Observatory partnership: the National Science Foundation (United States), National Research Council (Canada), Agencia Nacional de Investigaci\'{o}n y Desarrollo (Chile), Ministerio de Ciencia, Tecnolog\'{i}a e Innovaci\'{o}n (Argentina), Minist\'{e}rio da Ci\^{e}ncia, Tecnologia, Inova\c{c}\~{o}es e Comunica\c{c}\~{o}es (Brazil), and Korea Astronomy and Space Science Institute (Republic of Korea).

This research made use of matplotlib, a Python library for publication quality graphics \citep{Hunter:2007}; NumPy \citep{harris2020array}; Astroquery \citep{2019AJ....157...98G}; Astropy, a community-developed core Python package for Astronomy \citep{2018AJ....156..123A, 2013A&A...558A..33A}; and the VizieR catalogue access tool, CDS, Strasbourg, France.

This work has made use of data from the European Space Agency (ESA) mission {\it Gaia} (\url{https://www.cosmos.esa.int/gaia}), processed by the {\it Gaia} Data Processing and Analysis Consortium (DPAC, \url{https://www.cosmos.esa.int/web/gaia/dpac/consortium}). Funding for the DPAC has been provided by national institutions, in particular the institutions participating in the {\it Gaia} Multilateral Agreement.  Funding for the Sloan Digital Sky Survey IV has been provided by the Alfred P. Sloan Foundation, the U.S. Department of Energy Office of Science, and the Participating Institutions.

SDSS-IV acknowledges support and resources from the Center for High-Performance Computing at the University of Utah. The SDSS web site is www.sdss.org. SDSS-IV is managed by the Astrophysical Research Consortium for the Participating Institutions of the SDSS Collaboration including the Brazilian Participation Group, the Carnegie Institution for Science, Carnegie Mellon University, the Chilean Participation Group, the French Participation Group, Harvard-Smithsonian Center for Astrophysics, Instituto de Astrof\'isica de Canarias, The Johns Hopkins University, Kavli Institute for the Physics and Mathematics of the Universe (IPMU) / University of Tokyo, Lawrence Berkeley National Laboratory, Leibniz Institut f\"ur Astrophysik Potsdam (AIP), Max-Planck-Institut f\"ur Astronomie (MPIA Heidelberg), Max-Planck-Institut f\"ur Astrophysik (MPA Garching), Max-Planck-Institut f\"ur Extraterrestrische Physik (MPE), National Astronomical Observatories of China, New Mexico State University, New York University, University of Notre Dame, Observat\'ario Nacional / MCTI, The Ohio State University, Pennsylvania State University, Shanghai Astronomical Observatory, United Kingdom Participation Group, Universidad Nacional Aut\'onoma de M\'exico, University of Arizona, University of Colorado Boulder, University of Oxford, University of Portsmouth, University of Utah, University of Virginia, University of Washington, University of Wisconsin, Vanderbilt University, and Yale University. 
This publication makes use of data products from the Wide-field Infrared Survey Explorer \citep{2010AJ....140.1868W}, which is a joint project of the University of California, Los Angeles, and the Jet Propulsion Laboratory/California Institute of Technology, funded by the National Aeronautics and Space Administration. The Pan-STARRS1 Surveys (PS1) have been made possible through contributions of the Institute for Astronomy, the University of Hawaii, the Pan-STARRS Project Office, the Max-Planck Society and its participating institutes, the Max Planck Institute for Astronomy, Heidelberg and the Max Planck Institute for Extraterrestrial Physics, Garching, The Johns Hopkins University, Durham University, the University of Edinburgh, Queen's University Belfast, the Harvard-Smithsonian Center for Astrophysics, the Las Cumbres Observatory Global Telescope Network Incorporated, the National Central University of Taiwan, the Space Telescope Science Institute, the National Aeronautics and Space Administration under Grant No. NNX08AR22G issued through the Planetary Science Division of the NASA Science Mission Directorate, the National Science Foundation under Grant No. AST-1238877, the University of Maryland, and Eotvos Lorand University (ELTE).

\section*{Data Availability}
The ZTF data underlying this article are available at \url{https://irsa.ipac.caltech.edu/Missions/ztf.html}.
The other data underlying this article will be shared on reasonable request to the corresponding author.




\bibliographystyle{mnras}
\bibliography{references.bib,references2.bib} 



\appendix

\section{ZTF detection efficiency}\label{app:recovery}
\begin{figure}
    \centering
    \includegraphics{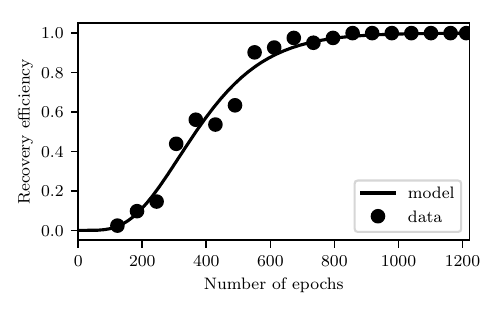}
    \caption{The period recovery efficiency as function of number of epochs. We used the ZTF lightcurve of \A. We randomly removed a fixed fraction of epochs and tested if our search algorithm recovered the known period. We fitted the curve with a simple probabilistic model (equation \ref{eq:eclipseprob}), and is a good approximation of the data. We use the model to rapidly evaluate in what fraction of the lightcurve an eclipse would be recovered.}
    \label{fig:recoveryefficiency}
\end{figure}

We determined the detection efficiency for \A\ as a function of the number of epochs, see Figure  \ref{fig:recoveryefficiency}. We did this by randomly removing a set fraction of points from the lightcurve, and tested if the BLS algorithm recovered the period. We model the curve with the function: 
\begin{equation}
P(k,p) = 1-\sum_{0}^{k} \binom{N}{k} p^{k} (1-p)^{N-k}
    \end{equation}\label{eq:eclipseprob}
    
with $k$ the number of in-eclipse points and $p$ the effective eclipse duty-cycle. For ZTF data of \A, the values of the best fit are $k=5$ and $p=0.0127$. This shows that the simple approximation is a good description of the actual recovery efficiency, and 5 in-eclipse points are enough to recover the orbital period.

\section{Follow-up observations}

Table \ref{tab:observ} shows an overview of all spectroscopic and photometric followup observations.

\begin{table*}
\small
 \centering
 \caption{Summary of the followup observations}
  \begin{tabular}{llclrrl}
  \hline\hline
Object & Date &    UT  &  Tele./Inst. & N$_{\rm exp}$ & Exp. time (s) & Wavelength \\
  \hline
\multicolumn{6}{c}{{\bf Photometry}}       \\
\A & 2019-05-29  & 06:37 - 08:09 & P200/CHIMERA & 1000 & \phantom{0}3.0 & $g$ \\
\A & 2019-05-29  & 06:37 - 08:09 & P200/CHIMERA & 1200 & \phantom{0}3.0 & $r$ \\
\A & 2019-05-30  & 04:03 - 05:52 & P200/CHIMERA & 1300 & \phantom{0}3.0 & $g$ \\
\A & 2019-05-30  & 04:03 - 05:52 & P200/CHIMERA & 1300 & \phantom{0}5.0 & $r$ \\
\A & 2019-06-29  & 04:27 - 11:30 & P200/CHIMERA & 4300 & \phantom{0}3.0 & $g$ \\
\A & 2019-06-29  & 04:27 - 11:30 & P200/CHIMERA & 4300 & \phantom{0}3.0 & $r$ \\
\A & 2019-09-06  & 20:42 - 21:54 & GTC/HiPERCAM & 335 & 12.9 & $u$ \\
\A & 2019-09-06  & 20:42 - 21:54 & GTC/HiPERCAM & 1340 & \phantom{0}3.2 & $g$ \\
\A & 2019-09-06  & 20:42 - 21:54 & GTC/HiPERCAM & 1340 & \phantom{0}3.2 & $r$ \\
\A & 2019-09-06  & 20:42 - 21:54 & GTC/HiPERCAM & 670 & \phantom{0}6.4 & $i$ \\
\A & 2019-09-06  & 20:42 - 21:54 & GTC/HiPERCAM & 335 & 12.9 & $z$ \\
\A & 2020-01-24  & 11:17 - 11:35 & P200/CHIMERA & 350 & \phantom{0}3.0 & $g$ \\
\A & 2020-01-24  & 11:17 - 11:35 & P200/CHIMERA & 175 & \phantom{0}6.0 & $r$ \\
\B & 2020-01-23  & 02:15 - 03:41 & P200/CHIMERA & 1700 & \phantom{0}3.0 & $g$ \\
\B & 2020-01-23  & 02:15 - 03:41 & P200/CHIMERA & 1600 & \phantom{0}3.0 & $r$ \\
\B & 2020-02-16  & 06:17 - 06:33 & Gemini/Alopeke & 320 & \phantom{0}3.0 & $g$ \\  
\B & 2020-02-16  & 06:17 - 06:33 & Gemini/Alopeke & 320 & \phantom{0}3.0 & $i$ \\
\B & 2020-02-17  & 05:25 - 05:41 & Gemini/Alopeke & 320 & \phantom{0}3.0 & $g$ \\  
\B & 2020-02-17  & 05:25 - 05:41 & Gemini/Alopeke & 320 & \phantom{0}3.0 & $i$ \\
\B & 2020-02-17  & 06:21 - 06:37 & Gemini/Alopeke & 320 & \phantom{0}3.0 & $g$ \\  
\B & 2020-02-17  & 06:21 - 06:37 & Gemini/Alopeke & 320 & \phantom{0}3.0 & $i$ \\
\B & 2020-02-18  & 06:21 - 06:37 & Gemini/Alopeke & 320 & \phantom{0}3.0 & $g$ \\  
\B & 2020-02-18  & 06:21 - 06:37 & Gemini/Alopeke & 320 & \phantom{0}3.0 & $i$ \\
\B & 2020-02-18  & 05:29 - 05:37 & Gemini/Alopeke & 320 & \phantom{0}3.0 & $g$ \\  
\B & 2020-02-18  & 05:29 - 05:37 & Gemini/Alopeke & 320 & \phantom{0}3.0 & $i$ \\

\C & 2020-10-17  & 02:12 - 06:58 & P200/CHIMERA & 3400 & \phantom{0}5.0 & $g$ \\
\C & 2020-10-17  & 02:12 - 06:58 & P200/CHIMERA & 3200 & \phantom{0}5.0 & $i$ \\
\C & 2020-10-21  & 02:52 - 08:00 & P200/CHIMERA & 2300 & \phantom{0}5.0 & $g$ \\
\C & 2020-10-21  & 02:49 - 08:00 & P200/CHIMERA & 2350 & \phantom{0}5.0 & $r$ \\

\D & 2020-01-23  & 03:49 - 06:07 & P200/CHIMERA & 2726 & \phantom{0}3.0 & $g$ \\   
\D & 2020-01-23  & 03:49 - 06:07 & P200/CHIMERA & 2720 & \phantom{0}3.0 & $r$ \\

\D & 2020-01-24  & 01:59 - 03:14 & P200/CHIMERA & 1500 & \phantom{0}3.0 & $g$ \\
\D & 2020-01-24  & 01:59 - 03:14 & P200/CHIMERA & 1500 & \phantom{0}3.0 & $r$ \\

\D & 2020-08-20  & 10:48 - 12:26 & P200/CHIMERA & 1900 & \phantom{0}3.0 & $g$ \\   
\D & 2020-08-20  & 10:49 - 12:29 & P200/CHIMERA & 1950 & \phantom{0}3.0 & $r$ \\

\D & 2020-08-27  & 11:29 - 11:44 & P200/CHIMERA & 150 & \phantom{0}6.0 & $g$ \\   
\D & 2020-08-27  & 11:29 - 11:44 & P200/CHIMERA & 150 & \phantom{0}6.0 & $r$ \\

\D & 2020-10-17  & 02:12 - 06:58 & P200/CHIMERA & 1500 & \phantom{0}5.0 & $g$ \\
\D & 2020-10-17  & 02:12 - 06:58 & P200/CHIMERA & 1500 & \phantom{0}5.0 & $i$ \\

\D & 2020-10-21  & 02:52 - 08:00 & P200/CHIMERA & 1000 & \phantom{0}5.0 & $g$ \\
\D & 2020-10-21  & 02:49 - 08:00 & P200/CHIMERA & 1000 & \phantom{0}5.0 & $r$ \\
\D & 2020-11-17  & 05:50 - 06:32 & P200/CHIMERA & 500 & \phantom{0}5.0 & $g$ \\
\D & 2020-11-17  & 05:50 - 06:36 & P200/CHIMERA & 550 & \phantom{0}5.0 & $r$ \\

\E & 2020-12-15  & 02:01 - 02:18 & P200/CHIMERA & 100 & 10.0 & $g$ \\
\E & 2020-12-15  & 02:01 - 02:18 & P200/CHIMERA & 100 & 10.0 & $i$ \\
\E & 2020-12-16 & 03:50 - 06:13 & P200/CHIMERA & 1650 & \phantom{0}5.0 & $g$ \\
\E & 2020-12-16  & 05:51 - 06:13 & P200/CHIMERA & 800 & 10.0 & $r$ \\

\noalign{\smallskip}
\multicolumn{6}{c}{{\bf Spectroscopy}}       \\
\A & 2019-05-07  &  11:00 - 12:45 &  Keck/LRIS/GR600 & 8 & 600 &  3200 - 5600\AA \\
\A & 2019-05-07  &  11:00 - 12:45 &  Keck/LRIS/R600 & 8 & 600 &  5600 - 8700\AA \\
\A & 2019-07-05  &  06:09 - 10:22 &  Keck/LRIS/GR600 & 47 & 300 &  3200 - 5600\AA \\
\A & 2019-07-05  &  06:09 - 10:22 &  Keck/LRIS/R600 & 41 & 300 &  5600 - 8700\AA \\

\B & 2019-12-03  &  10:33 - 10:43 &  Keck/LRIS/GR600 & 1 & 600 &  3200 - 5600\AA \\
\B & 2019-12-03  &  10:33 - 10:43 &  Keck/LRIS/R600 & 1 & 600 &  5600 - 8700\AA \\

\B & 2020-01-25  &  05:22 - 07:30 &  Keck/LRIS/GR600 & 20 & 300 &  3200 - 5600\AA \\
\B & 2020-01-25  &  05:22 - 07:30 &  Keck/LRIS/R600 & 23 & 300 &  5600 - 8700\AA \\

\D & 2020-01-25  &  07:47 - 09:17 &  Keck/LRIS/GR600 & 20 & 200 &  3200 - 5600\AA \\
\D & 2020-01-25  &  07:47 - 09:17 &  Keck/LRIS/R600 & 23 & 200 &  5600 - 8700\AA \\

\C & 2020-09-16  &  12:24 - 13:10 &  Keck/LRIS/GR600 & 11 & 220 &  3200 - 5600\AA \\
\C & 2020-09-16  &  12:25 - 13:10 &  Keck/LRIS/R600 & 8 & 220 &  5600 - 8700\AA \\

\E & 2020-12-10  &  08:19 - 08:24 &  Keck/LRIS/GR600 & 1 & 300 & 3200 - 5600\AA \\
\E & 2020-12-10  &  08:19 - 08:24 &  Keck/LRIS/R300 & 1 & 300 &  5600 - 8700\AA \\
\E & 2020-12-17  &  07:29 - 08:36 &  Keck/LRIS/GR600 & 12 & 300 &  3200 - 5600\AA \\
\E & 2020-12-17  &  07:29 - 08:36 &  Keck/LRIS/R600 & 8 & 420 &  5600 - 8700\AA \\
   \hline
\end{tabular}
\label{tab:observ}
\end{table*}

\section{Lightcurve and binary parameters}
Tables \ref{tab:lcpars_A}--\ref{tab:lcpars_D} show the lightcurve model parameters and the derived binary parameter for each object and each band.

\begin{table*} 
\centering
    \renewcommand{\arraystretch}{1.25}
\begin{tabular}{l|lllll}
& $u$ & $g$ & $r$ & $i$ & $z$ \\
\hline
$q$ & $ 0.026^{+0.010}_{-0.009}$ & $0.025^{+0.008 }_{-0.006 }$ & $0.024^{+0.012 }_{-0.009 }$ & $0.025^{+0.012 }_{-0.010 }$ & $0.026^{+0.013 }_{-0.011 } $\\
$i$ ($^{\circ}$) & $ 82.7^{+0.9 }_{-0.8 }$ & $82.8^{+0.6 }_{-0.6 }$ & $82.9^{+1.0 }_{-1.0 }$ & $82.8^{+1.1 }_{-1.0 }$ & $82.7^{+1.1 }_{-1.0 } $\\
$r_1$ & $ 0.020^{+0.003 }_{-0.002 }$ & $0.019^{+0.001 }_{-0.002 }$ & $0.020^{+0.003 }_{-0.002 }$ & $0.019^{+0.003 }_{-0.002 }$ & $0.021^{+0.004 }_{-0.003 } $\\
velocity scale ($\mathrm{km s^{-1}}$) & $ 583^{+16 }_{-20 }$ & $593^{+13 }_{-13 }$ & $585^{+18 }_{-20 }$ & $588^{+18 }_{-24 }$ & $575^{+23 }_{-27 } $\\
$t_0$ (s) & $ -1.4^{+0.6 }_{-0.7 }$ & $-0.5^{+0.2 }_{-0.2 }$ & $-0.7^{+0.2 }_{-0.2 }$ & $-1.2^{+0.3 }_{-0.4 }$ & $-0.8^{+0.8 }_{-0.8 } $\\
\hline
$a$ (\Rsun)  & $0.492^{0.014}_{-0.017}$ &  $0.501^{0.010}_{-0.011}$ &  $0.494^{0.015}_{-0.017}$ & $0.496^{0.015}_{-0.019}$ &  $0.485^{0.020}_{-0.022}$\\
$M_1$ (\Msun)& $ 0.85^{+0.07 }_{-0.08 }$ & $0.90^{+0.05 }_{-0.05 }$ & $0.87^{+0.07 }_{-0.08 }$ & $0.88^{+0.07 }_{-0.09 }$ & $0.82^{+0.09 }_{-0.10 } $\\
$M_2$ (\Msun)& $ 0.022^{+0.011 }_{-0.009 }$ & $0.023^{+0.009 }_{-0.006 }$ & $0.021^{+0.013 }_{-0.009 }$ & $0.022^{+0.013 }_{-0.010 }$ & $0.021^{+0.014 }_{-0.011 } $\\
$R_1$ (\Rsun)& $ 0.010^{+0.001 }_{-0.001 }$ & $0.009^{+0.001 }_{-0.001 }$ & $0.010^{+0.001 }_{-0.001 }$ & $0.010^{+0.001 }_{-0.001 }$ & $0.010^{+0.001 }_{-0.001 } $\\
$R_2$ (\Rsun)& $ 0.068^{+0.009 }_{-0.011 }$ & $0.068^{+0.007 }_{-0.007 }$ & $0.066^{+0.011 }_{-0.011 }$ & $0.068^{+0.011 }_{-0.012 }$ & $0.067^{+0.012 }_{-0.013 } $\\
$K_1$ ($\mathrm{km s^{-1}}$) & $ 14^{+6 }_{-5 }$ & $14^{+5 }_{-4 }$ & $14^{+7 }_{-5 }$ & $14^{+7 }_{-6 }$ & $14^{+8 }_{-6 } $\\
$K_2$ ($\mathrm{km s^{-1}}$) & $ 563^{+11 }_{-15 }$ & $573^{+7 }_{-9 }$ & $566^{+10 }_{-13 }$ & $569^{+10 }_{-16 }$ & $555^{+14 }_{-20 } $\\
$\log g_1$ & $ 8.39^{+0.10 }_{-0.12 }$ & $8.46^{+0.08 }_{-0.07 }$ & $8.40^{+0.11 }_{-0.12 }$ & $8.43^{+0.11 }_{-0.14 }$ & $8.33^{+0.13 }_{-0.15 } $\\
$\log g_2$ & $ 4.81^{+0.08 }_{-0.10 }$ & $4.88^{+0.07 }_{-0.06 }$ & $4.91^{+0.10 }_{-0.11 }$ & $4.96^{+0.09 }_{-0.12 }$ & $4.96^{+0.10 }_{-0.12 } $\\
\end{tabular}
\caption{The best-fit parameters of the \textit{lcurve} models of the lightcurves of \A\ for each band. The reported values are the median with the 68 percentile interval. The $t_0$ is given as the deviation from the ephemeris in seconds. The parameters above the horizontal line are model variables, below the line are derived parameters.}
\label{tab:lcpars_A}
\end{table*}

\begin{table*}
    \centering
    \renewcommand{\arraystretch}{1.25}
    \begin{tabular}{l|lll}
& $g$ & $r$ & $i$ \\
\hline
$q$  &  $0.021^{0.016}_{-0.007}$ &  $0.016^{0.020}_{-0.004}$ &  $0.022^{0.017}_{-0.009}$\\
$i$ ($^{\circ}$) &  $86.5^{2.0}_{-2.1}$ &  $87.4^{2.1}_{-3.1}$ &  $86.3^{2.6}_{-2.3}$\\
$r_1$  &  $0.024^{0.004}_{-0.005}$ &  $0.018^{0.007}_{-0.006}$ &  $0.024^{0.005}_{-0.005}$\\
velocity scale ($\mathrm{km s^{-1}}$)  &  $586^{34}_{-28}$ &  $624^{41}_{-53}$ &  $586^{39}_{-35}$\\
$t_0$ (s) &  $0.08^{0.50}_{-0.54}$ &  $-0.3^{1.2}_{-1.2}$ &  $0.7^{3.4}_{-3.9}$\\
$T_\mathrm{disc}$ (K)  &  $1900^{1200}_{-1300}$ &  $2900.0^{1000}_{-1600}$ &  $2000.0^{1200}_{-1500}$\\
$r_\mathrm{spot}$  &  $0.49^{0.19}_{-0.18}$ &  $0.40^{0.25}_{-0.11}$ &  $0.46^{0.23}_{-0.18}$\\
$l_\mathrm{spot}$  &  $0.043^{0.038}_{-0.033}$ &  $0.042^{0.038}_{-0.032}$ &  $0.043^{0.045}_{-0.035}$\\
$T_\mathrm{spot}$ (K)  &  $2900.0^{2500.0}_{-1900.0}$ &  $3300.0^{4600.0}_{-2400.0}$ &  $3100.0^{2500.0}_{-2300.0}$\\
\hline
$a$ (\Rsun) &  $0.446^{0.026}_{-0.021}$ &  $0.480^{0.029}_{-0.04}$ &  $0.380^{0.058}_{-0.051}$\\
$M_1$ (\Msun) &  $0.79^{0.13}_{-0.10}$ &  $0.95^{0.20}_{-0.21}$ &  $0.78^{0.16}_{-0.13}$\\
$M_2$ (\Msun) &  $0.0165^{0.0179}_{-0.0064}$ &  $0.015^{0.0239}_{-0.0054}$ &  $0.0169^{0.0185}_{-0.0086}$\\
$R_1$ (\Rsun) &  $0.0106^{0.0012}_{-0.0015}$ &  $0.0087^{0.0025}_{-0.0022}$ &  $0.0106^{0.0015}_{-0.0017}$\\
$R_2$ (\Rsun) &  $0.058^{0.015}_{-0.008}$ &  $0.056^{0.019}_{-0.007}$ &  $0.045^{0.022}_{-0.009}$\\
$K_1$ ($\mathrm{km s^{-1}}$)  &  $12.1^{10.2}_{-4.1}$ &  $9.5^{13.1}_{-2.5}$ &  $12.3^{10.4}_{-5.5}$\\
$K_2$ ($\mathrm{km s^{-1}}$)  &  $573^{22}_{-24}$ &  $608^{37}_{-46}$ &  $571.0^{29}_{-29}$\\
$\log g_1$  &  $8.30^{0.20}_{-0.15}$ &  $8.55^{0.34}_{-0.33}$ &  $8.30^{0.24}_{-0.18}$\\
$\log g_2$  &  $4.78^{0.11}_{-0.07}$ &  $4.87^{0.16}_{-0.09}$ &  $4.88^{0.12}_{-0.10}$\\
    \end{tabular}
    \caption{The best-fit parameters of the \textit{lcurve} models of the lightcurves of \E, similar to Table \ref{tab:lcpars_A}.}
    \label{tab:lcpars_E}
\end{table*}

\begin{table}
    \centering
    \renewcommand{\arraystretch}{1.25}
    \begin{tabular}{l|lll}
& $g$ & $r$ & $i$ \\
\hline
$q$  &  $0.017^{0.007}_{-0.004}$ &  $0.013^{0.014}_{-0.006}$ &  $0.0051^{0.0018}_{-0.0003}$ \\
$i$ ($^{\circ}$) &  $85.3^{0.8}_{-0.9}$ &  $86.1^{2.2}_{-2.0}$ &  $89.3^{0.5}_{-1.5}$ \\
$r_1$  &  $0.0227^{0.0027}_{-0.0024}$ &  $0.0249^{0.0085}_{-0.0059}$ &  $0.0302^{0.0023}_{-0.0035}$ \\
velocity scale ($\mathrm{km s^{-1}}$)  &  $604^{19}_{-20}$ &  $588^{44}_{-54}$ &  $554^{23}_{-15}$ \\
$t_0$ (s) & $-0.37^{0.14}_{-0.15}$ &  $0.73^{0.69}_{-0.68}$ &  $0.70^{0.38}_{-0.38}$\\ 
$T_\mathrm{disc}$ (K) &  $3330^{130}_{-130}$ &  $2780^{590}_{-790}$ &  $3280^{790}_{-640}$ \\
$r_\mathrm{spot}$  &  $0.675^{0.075}_{-0.081}$ &  $0.44^{0.23}_{-0.11}$ &  $0.58^{0.15}_{-0.11}$ \\
$l_\mathrm{spot}$  &  $0.044^{0.039}_{-0.033}$ &  $0.041^{0.039}_{-0.032}$ &  $0.033^{0.044}_{-0.026}$ \\
$T_\mathrm{spot}$ (K) &  $2000^{1400}_{-1300}$ &  $4800^{3600}_{-3000}$ &  $6300^{11900}_{-4300}$ \\
\hline
$a$ (\Rsun) &  $0.443^{0.014}_{-0.015}$ &  $0.431^{0.032}_{-0.04}$ &  $0.405^{0.017}_{-0.011}$ \\
$M_1$ (\Msun) &  $0.83^{0.07}_{-0.08}$ &  $0.77^{0.17}_{-0.19}$ &  $0.65^{0.08}_{-0.05}$ \\
$M_2$ (\Msun) &  $0.0144^{0.007}_{-0.005}$ &  $0.010^{0.015}_{-0.006}$ &  $0.0032^{0.0017}_{-0.0003}$ \\
$R_1$ (\Rsun) &  $0.0100^{0.0008}_{-0.0008}$ &  $0.0107^{0.0023}_{-0.0019}$ &  $0.0122^{0.0006}_{-0.0010}$ \\
$R_2$ (\Rsun) &  $0.0537^{0.0076}_{-0.0063}$ &  $0.048^{0.017}_{-0.013}$ &  $0.0331^{0.0049}_{-0.0011}$ \\
$K_1$ ($\mathrm{km s^{-1}}$)  &  $10.3^{4.3}_{-2.8}$ &  $7.5^{9.0}_{-4.1}$ &  $2.76^{1.15}_{-0.17}$ \\
$K_2$ ($\mathrm{km s^{-1}}$)  &  $592^{14}_{-17}$ &  $579.0^{34}_{-49}$ &  $551.0^{22}_{-14}$ \\
$\log g_1$  &  $8.35^{0.10}_{-0.11}$ &  $8.27^{0.26}_{-0.29}$ &  $8.07^{0.13}_{-0.08}$ \\
$\log g_2$  &  $4.77^{0.06}_{-0.06}$ &  $4.79^{0.14}_{-0.14}$ &  $4.66^{0.06}_{-0.02}$ \\
    \end{tabular}
    \caption{The best-fit parameters of the \textit{lcurve} models of the lightcurves of \B, similar to Table \ref{tab:lcpars_A}.}
    \label{tab:lcpars_B}
\end{table}

\begin{table*}
    \centering
    \renewcommand{\arraystretch}{1.25}
    \begin{tabular}{l|lll}
& $g$ & $r$ & $i$ \\
\hline
$q$ & $ 0.034^{+0.010 }_{-0.002 }$ & $0.033^{+0.004 }_{-0.002 }$ & $0.034^{+0.007 }_{-0.003 } $\\
$i$ ($^{\circ}$)& $ 87.0^{+0.5 }_{-1.5 }$ & $87.2^{+0.6 }_{-0.7 }$ & $87.0^{+0.6 }_{-1.1 } $\\
$r_1$ & $ 0.030^{+0.002 }_{-0.003 }$ & $0.030^{+0.002 }_{-0.002 }$ & $0.027^{+0.003 }_{-0.003 } $\\
velocity scale ($\mathrm{km s^{-1}}$) & $ 675^{+18 }_{-11 }$ & $675^{+12 }_{-11 }$ & $696^{+18 }_{-21 } $\\
$t_0$ (s) & $1.52^{0.14}_{-0.21}$ &  $1.27^{0.29}_{-0.37}$ &  $1.79^{0.62}_{-0.71}$\\
$T_\mathrm{disc}$ (K) & $ 5300^{+260 }_{-370 }$ & $5490^{+330 }_{-300 }$ & $5000^{+550 }_{-530 } $\\
texp disc & $ -0.74^{+0.23 }_{-0.20 }$ & $-0.22^{+0.15 }_{-0.21 }$ & $-0.72^{+0.23 }_{-0.18 } $\\
$r_\mathrm{spot}$ & $ 0.55^{+0.04 }_{-0.03 }$ & $0.49^{+0.01 }_{-0.02 }$ & $0.49^{+0.02 }_{-0.02 } $\\
$l_\mathrm{spot}$ & $ 0.07^{+0.02 }_{-0.02 }$ & $0.06^{+0.02 }_{-0.03 }$ & $0.03^{+0.02 }_{-0.02 } $\\
angle spot ($^{\circ}$)& $ 153^{+14 }_{-41 }$ & $105^{+23 }_{-13 }$ & $111^{+42 }_{-126 } $\\
yaw spot ($^{\circ}$)& $ -118^{+54 }_{-26 }$ & $-34^{+36 }_{-31 }$ & $-89^{+128 }_{-42 } $\\
$T_\mathrm{spot}$ (K)  & $ 11100^{+2000 }_{-1700 }$ & $13200^{+4200 }_{-2400 }$ & $10900^{+3700 }_{-2400 } $\\
cfrac spot & $ 0.97^{+0.01 }_{-0.02 }$ & $0.99^{+0.00 }_{-0.01 }$ & $0.92^{+0.04 }_{-0.07 } $\\
\hline
$a$ (\Rsun) & $0.347^{0.009}_{-0.006}$ &  $0.347^{0.006}_{-0.006}$ &  $0.357^{0.009}_{-0.010}$\\
$M_1$ (\Msun)& $ 0.76^{+0.05 }_{-0.04 }$ & $0.76^{+0.04 }_{-0.04 }$ & $0.83^{+0.06 }_{-0.07 } $\\
$M_2$ (\Msun)& $ 0.026^{+0.010 }_{-0.002 }$ & $0.025^{+0.004 }_{-0.003 }$ & $0.028^{+0.007 }_{-0.004 } $\\
$R_1$ (\Rsun)& $ 0.010^{+0.000 }_{-0.001 }$ & $0.010^{+0.000 }_{-0.000 }$ & $0.009^{+0.001 }_{-0.001 } $\\
$R_2$ (\Rsun)& $ 0.049^{+0.006 }_{-0.002 }$ & $0.048^{+0.002 }_{-0.002 }$ & $0.051^{+0.004 }_{-0.002 } $\\
$K_1$ ($\mathrm{km s^{-1}}$) & $ 22^{+7 }_{-2 }$ & $21^{+3 }_{-2 }$ & $23^{+5 }_{-2 } $\\
$K_2$ ($\mathrm{km s^{-1}}$) & $ 651^{+13 }_{-10 }$ & $652^{+10 }_{-10 }$ & $672^{+15 }_{-18 } $\\
$\log g_1$ & $ 8.32^{+0.09 }_{-0.05 }$ & $8.33^{+0.06 }_{-0.05 }$ & $8.45^{+0.10 }_{-0.11 } $\\
$\log g_2$ & $5.29^{0.07}_{-0.02}$ &  $5.33^{0.03}_{-0.02}$ & $5.39^{0.04}_{-0.03}$\\
    \end{tabular}
    \caption{The best-fit parameters of the \textit{lcurve} models of the lightcurves of \C, similar to Table \ref{tab:lcpars_A}.}
    \label{tab:lcpars_C}
\end{table*}

\begin{table*}
    \centering
    \renewcommand{\arraystretch}{1.25}
    \begin{tabular}{l|lll}
& $g$ & $r$ & $i$ \\
\hline
$q$ & $ 0.024^{+0.003 }_{-0.004 }$ & $0.033^{+0.006 }_{-0.005 }$ & $0.024^{+0.008 }_{-0.005 } $\\
$i$ ($^{\circ}$)& $ 86.5^{+0.8 }_{-0.5 }$ & $85.0^{+0.7 }_{-0.7 }$ & $86.4^{+1.1 }_{-1.2 } $\\
$r_1$ & $ 0.032^{+0.002 }_{-0.002 }$ & $0.027^{+0.003 }_{-0.002 }$ & $0.032^{+0.004 }_{-0.005 } $\\
velocity scale ($\mathrm{km s^{-1}}$) & $ 683^{+13 }_{-14 }$ & $712^{+16 }_{-19 }$ & $678^{+30 }_{-25 } $\\
$t_0$ (s)  & $1.52^{0.14}_{-0.21}$ &  $1.27^{0.29}_{-0.37}$ &  $1.79^{0.62}_{-0.71}$ \\
$r_\mathrm{disc}$ & $ 0.60^{+0.08 }_{-0.07 }$ & $0.59^{+0.11 }_{-0.13 }$ & $0.57^{+0.12 }_{-0.10 } $\\
$T_\mathrm{disc}$ (K) & $ 4280^{+330 }_{-240 }$ & $3290^{+320 }_{-220 }$ & $3940^{+620 }_{-450} $\\
$r_\mathrm{spot}$ & $ 0.50^{+0.03 }_{-0.02 }$ & $0.44^{+0.03 }_{-0.03 }$ & $0.46^{+0.10 }_{-0.10 } $\\
$l_\mathrm{spot}$ & $ 0.02^{+0.01 }_{-0.01 }$ & $0.03^{+0.03 }_{-0.01 }$ & $0.06^{+0.03 }_{-0.02 } $\\
angle spot ($^{\circ}$) & $ 62^{+91 }_{-11 }$ & $57^{+38 }_{-20 }$ & $-49^{+116 }_{-64 } $\\
yaw spot ($^{\circ}$) & $ -27^{+11 }_{-94 }$ & $-11^{+20 }_{-40 }$ & $71^{+64 }_{-108 } $\\
$T_\mathrm{spot}$ (K) & $ 14800^{+3200 }_{-2500 }$ & $10000^{+3000 }_{-2100 }$ & $13800^{+4500 }_{-3700 } $\\
cfrac spot & $ 0.94^{+0.02 }_{-0.01 }$ & $0.93^{+0.01 }_{-0.02 }$ & $0.95^{+0.03 }_{-0.04 } $\\
\hline
$a$ (\Rsun) &  $0.330^{0.006}_{-0.007}$ &  $0.346^{0.008}_{-0.009}$ &  $0.330^{0.014}_{-0.012}$\\
$M_1$ (\Msun)& $ 0.79^{+0.04 }_{-0.05 }$ & $0.89^{+0.06 }_{-0.07 }$ & $0.78^{+0.10 }_{-0.08 } $\\
$M_2$ (\Msun)& $ 0.019^{+0.003 }_{-0.004 }$ & $0.030^{+0.007 }_{-0.006 }$ & $0.019^{+0.009 }_{-0.005 } $\\
$R_1$ (\Rsun)& $ 0.010^{+0.001 }_{-0.000 }$ & $0.009^{+0.001 }_{-0.001 }$ & $0.011^{+0.001 }_{-0.001 } $\\
$R_2$ (\Rsun)& $ 0.044^{+0.002 }_{-0.003 }$ & $0.051^{+0.004 }_{-0.003 }$ & $0.044^{+0.006 }_{-0.004 } $\\
$K_1$ ($\mathrm{km s^{-1}}$) & $ 16^{+2 }_{-3 }$ & $23^{+4 }_{-4 }$ & $16^{+6 }_{-4 } $\\
$K_2$ ($\mathrm{km s^{-1}}$) & $ 666^{+11 }_{-11 }$ & $686^{+12 }_{-15 }$ & $661^{+24 }_{-22 } $\\
$\log g_1$ & $ 8.15^{+0.06 }_{-0.06 }$ & $8.30^{+0.08 }_{-0.09 }$ & $8.12^{+0.15 }_{-0.12 } $\\
$\log g_2$ & $ 4.96^{+0.02 }_{-0.03 }$ & $5.06^{+0.03 }_{-0.03 }$ & $5.03^{+0.05 }_{-0.04 } $\\
    \end{tabular}
    \caption{The best-fit parameters of the \textit{lcurve} models of the lightcurves of \D, similar to Table \ref{tab:lcpars_A}.}
    \label{tab:lcpars_D}
\end{table*}

\bsp	
\label{lastpage}
\end{document}